\documentclass[usenatbib]{mn2e}

\pdfoutput=1

\usepackage[pdftex]{graphicx} 

\usepackage{times}

\usepackage{rotate} 

\usepackage{amssymb} 

\usepackage{rotating}

\usepackage{epstopdf} 

\usepackage{graphicx, subfigure}

\usepackage{array}

\title[The dependence of the X-ray AGN clustering on the properties of the host galaxy]{The dependence of the X-ray AGN clustering on the properties of the host galaxy.}

\author[G. Mountrichas, A. Georgakakis and I. Georgantopoulos]{G. Mountrichas $^1$, A. Georgakakis$^1$ and I. Georgantopoulos$^1$ \\ \\ 
$^1$National Observatory of Athens, V.  Paulou  \& I.  Metaxa, 15236,  Greece}

\begin{document}
\maketitle
\label{firstpage}

\begin{abstract}  
We study the clustering dependence of X-ray AGN and normal galaxies on the stellar mass (M$_\star$), star-formation rate (SFR) and specific star-formation rate (sSFR) of the (host) galaxy. Towards this end, we use 407 X-ray AGN from the XMM-XXL survey ($\sim$ 25\,deg$^2$ in the North) and $\sim45,000$ galaxies in the VIPERS field (W1: $\sim$16\,deg$^2$). We match the AGN and galaxy samples to have the same M$_\star$, SFR and redshift distributions. Based on our results, the two populations live in DMHs with similar mass ($\log M / (M_{\odot} \,  h^{-1})\approx 12.8$) and have similar dependence on the galaxy properties. Specifically, our measurements show a positive dependence of the AGN and galaxy clustering on M$_\star$ and a negative dependence on SFR and sSFR. We also find that the X-ray clustering is independent of the location of the host galaxy above or below the star-forming main sequence.
\end{abstract}


\begin{table*}
\caption{The models and the values for their free parameters used by CIGALE for the SED fitting of our X-ray AGN and the galaxy samples. The Fritz et al. (2006) model for the AGN emission has been used only for the X-ray sample. $\tau$ is the e-folding time of the main stellar population model in M\rm{yr}, age is the age of the main stellar population in the galaxy in M\rm{yr} (the precision is 1\,M\rm{yr}), burst age is the age of the late burst in M\rm{yr} (the precision is 1 M\rm{yr}). $\beta$ and $\gamma$ are the parameters used to define the law for the spatial behaviour of density of the torus density. The latter has the functional form $\rho (r, \theta) \propto  r^\beta e^{-\gamma | cos \theta |}$, where r is the radial distance and $\theta$ is the polar distance. $\Theta$ and $\Psi$ are the opening and viewing angles of the torus, respectively, with $\Psi=0$ for Type-2 AGN and  $\Psi=90$ for Type-1 AGN. The AGN fraction is measured as the AGN emission relative to infrared luminosity ($1-1000\,\mu m$).} 
\centering
\setlength{\tabcolsep}{1.5mm}
\begin{tabular}{cc}
       \hline
Parameter &  Model/values \\
	\hline
\multicolumn{2}{c}{Stellar population synthesis model} \\
\\
Initial Mass Function & Salpeter\\
Metallicity &  0.0001, 0.008, 0.02, 0.05 (Solar) \\
Single Stellar Population Library & Bruzual \& Charlot (2003) \\
\hline
\multicolumn{2}{c}{double-exponentially-decreasing (2$\tau$-dec) model} \\
$\tau$ & 100, 500, 1000, 3000, 5000, 10000, 20000 \\ 
age & 500, 1000, 2000, 3000, 4000, 5000, 6000, 7000, 8000, 9000, 10000, 11000, 12000, 13000 \\
burst age & 100, 200, 400, 500 \\
\hline
\multicolumn{2}{c}{Dust extinction} \\
Dust attenuation law & Calzetti et al. (2000) \\
Reddening E(B-V) & 0.05, 0.1, 0.15, 0.2, 0.25, 0.3, 0.35, 0.4, 0.5, 0.6, 0.8, 1.2 \\ 
E(B-V) reduction factor between old and young stellar population & 0.44 \\
\hline
\multicolumn{2}{c}{Fritz et al. (2006) model for AGN emission} \\ 
Ratio between outer and inner dust torus radii &   10, 60, 150 \\
9.7 $\mu m$ equatorial optical depth & 0.1, 0.3, 1.0, 2.0, 6.0, 10.0 \\
$\beta$ & -0.5 \\
$\gamma$ & 0.0, 2.0, 6.0 \\
$\Theta$ & 100 \\
$\Psi$ & 0.001, 10.100, 20.100, 30.100, 40.100, 50.100, 60.100, 70.100, 80.100, 89.990 \\
AGN fraction & 0.1, 0.2, 0.3, 0.5, 0.6, 0.8 \\
\hline
\label{table_cigale}
\end{tabular}
\end{table*}

\section{Introduction}

In the last decade there has been growing evidence supporting the coeval growth of the galaxies and their resident Supermassive Black Hole \citep[SMBH, e.g.][]{Boyle1998, Kormendy2013}. The mass of the SMBH is also correlated with the properties of its bulge, parametrised by the luminosity \citep{Magorrian2008}, or the velocity dispersion \citep{Ferrarese2000}. Apart from the observational evidence, theoretical and semi-analytical models, also assume a connection between the active SMBH (AGN) and its host galaxy, through mergers \citep[e.g.][]{Hopkins2006a, Hopkins2008a}. 

However, it is still not clear what are the physical mechanisms that drive the BH growth, how the large-scale environment (Dark Matter Halo mass, DMH mass) affects these feeding mechanisms and what is the connection between the properties of BHs (e.g. BH mass, Eddington ratio) and the host galaxy properties, e.g. Star-Formation Rate (SFR), stellar mass (M$_\star$) and specific SFR (sSFR).

One way to shed light on these questions is via a clustering analysis \citep[e.g.][]{Croom2004, Coil2009, Mountrichas2009, Allevato2012, Koutoulidis2013, Powell2018}. Clustering allows us to study the environment that extragalactic sources live in \citep[e.g.][]{Miyaji2011, Cappelluti2012, Krumpe2018}. Thus, comparing the environment of AGN that have different SBMH characteristics and/or live in galaxies with different properties can put constraints on how AGN and galaxies interact and co-evolve \citep[e.g.][]{Krumpe2015}. Moreover, comparison of the observational measurements with the predictions of theoretical models can shed light on the feeding mechanism(s) that trigger the AGN activity \citep[e.g.][]{Bonoli2009, Marulli2009, Fanidakis2013a}.

Initially, many clustering works focused on studying the correlation of the AGN clustering on X-ray luminosity, L$_X$, since L$_X$ is a proxy of the AGN power {\citep[e.g.][]{Lusso2012}. The results of these studies though are controversial. Some studies claim a weak dependence of X-ray clustering on L$_X$ \citep[e.g.][]{Starikova2011, Koutoulidis2013}, other find no statistical significant dependence \citep[e.g.][]{Coil2009, Mountrichas2012} while some works suggest a negative dependence of the AGN clustering on luminosity \citep{Mountrichas2016}.

More recent studies have suggested that the apparent discrepancies on the luminosity dependence of the X-ray clustering could be incited by other properties of the active SuperMassive Black Hole (SMBH) and/or of the host galaxy that are closely related to the physical processes of galaxies. It has been found that, at least in the case of broad-line, luminous AGN in the local universe, there is a weak clustering dependence on BH mass, whereas no statistically significant dependence was detected on accretion rate \citep{Krumpe2015}. However, the clustering dependence of X-ray AGN on host galaxy properties has not been directly studied. \cite{Georgakakis2014} found that X-ray AGN, star-forming and passive galaxies live in DMHs with similar mass and attributed this to the similar stellar mass distributions of the three populations. \cite{Mendez2016} studied the clustering of X-ray, radio and infrared (IR) AGN using the PRIMUS and DEEP2 redshift surveys. They claim that the observed differences in the clustering of AGN selected at different wavelengths can be attributed to the different distributions in both stellar mass and star formation rate of their host populations. 

{The scope of this work is to directly study the X-ray AGN clustering dependence on SFR, M$_{\star}$ and sSFR in an attempt to disentangle those three galaxy properties and compare the clustering properties of active and normal galaxies. Towards this end, we use X-ray AGN from one of the largest contiguous fields, the XMM-XXL area \citep{Pierre2016} and cross-correlate them with galaxies in the VIPERS field \citep{Guzzo2014, Garilli2014}. We also match our AGN and galaxy samples to have the same stellar mass, SFR and redshift distributions and compare their clustering properties. We study the dependence of the galaxy clustering on M$_{\star}$, SFR and sSFR and compare our findings with those for AGN to see whether the existence of an active SMBH affects the clustering properties of the host galaxy. Throughout this paper we use $\Omega_m=0.3$  $\Omega_{\Lambda}=0.7$  and  $\sigma_8=0.8$. For the clustering analysis the Hubble constant is set to $H_{0}=100$\,km\,s$^{-1}$Mpc$^{-1}$ and all relevant quantities, such halo masses, are  parametrised by $h=H_0/100$. In the calculation of the X-ray luminosities we fix  $H_0=70$\,km\,s$^{-1}$\,Mpc$^{-1}$
(i.e. $h=0.7$). This is to allow comparison with previous studies that also follow similar conventions.

\begin{figure*}
\centering
\begin{subfigure}
  \centering
  \includegraphics[width=.32\linewidth]{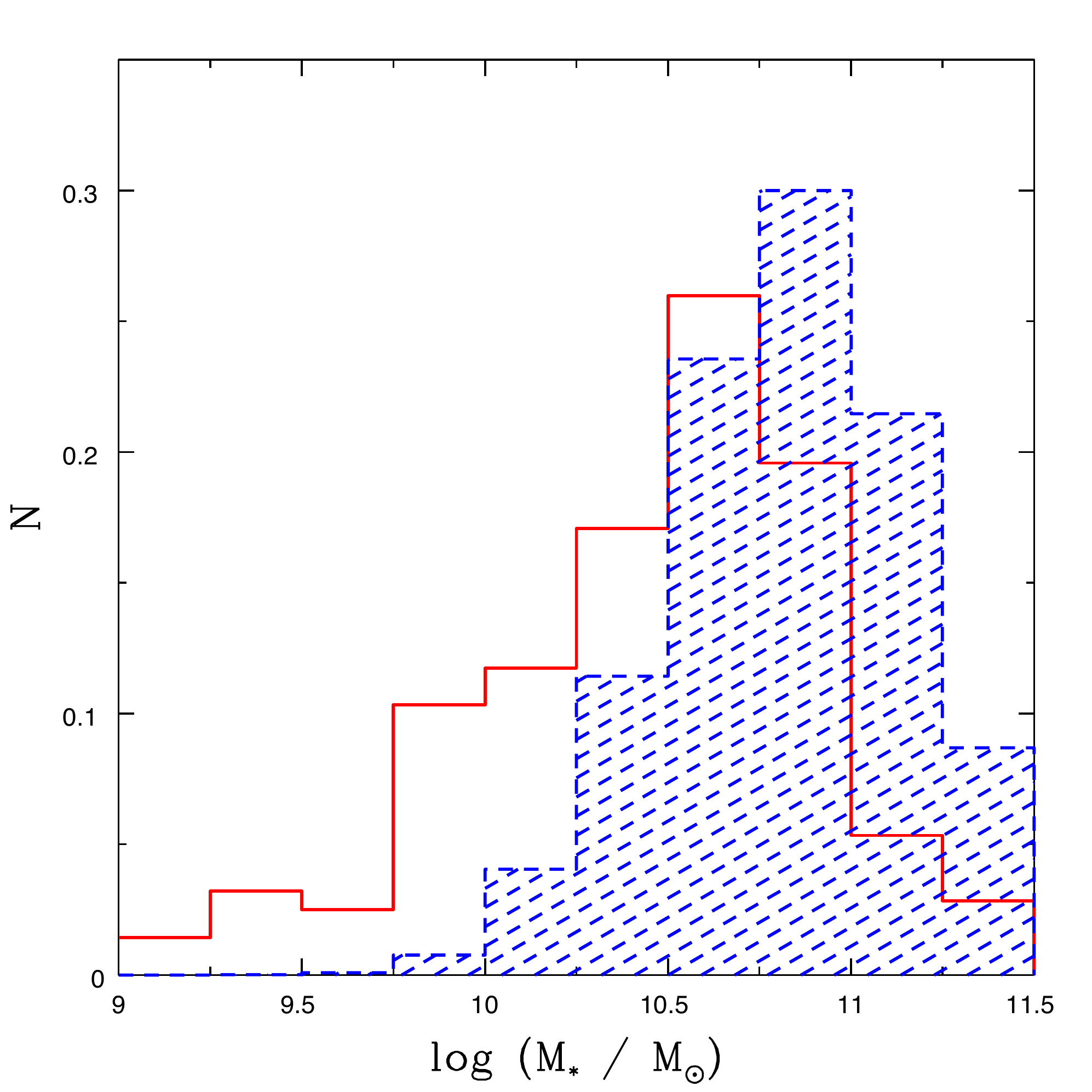}
\end{subfigure}
\begin{subfigure}
  \centering
  \includegraphics[width=.32\linewidth]{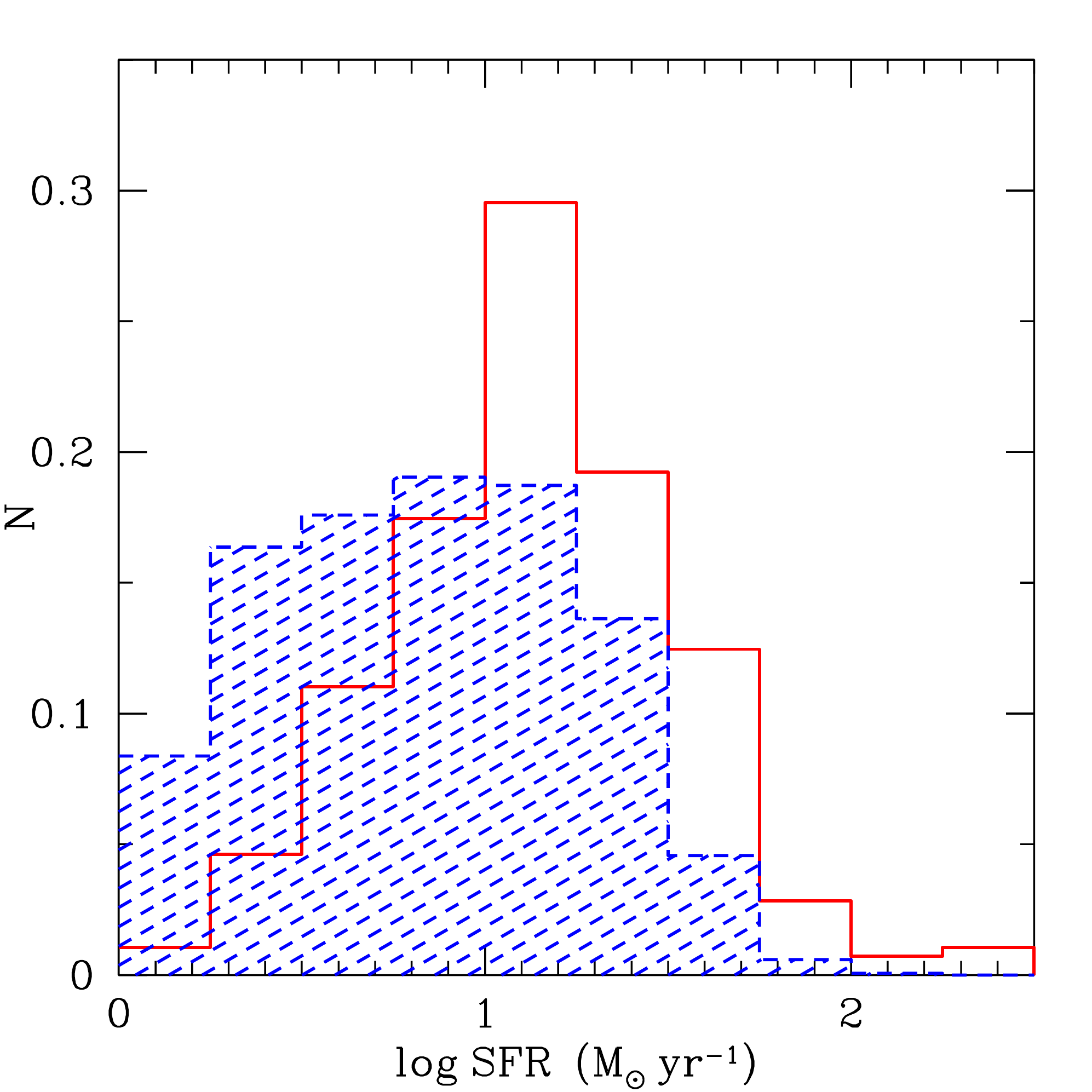}
\end{subfigure}
\begin{subfigure}
  \centering
  \includegraphics[width=.32\linewidth]{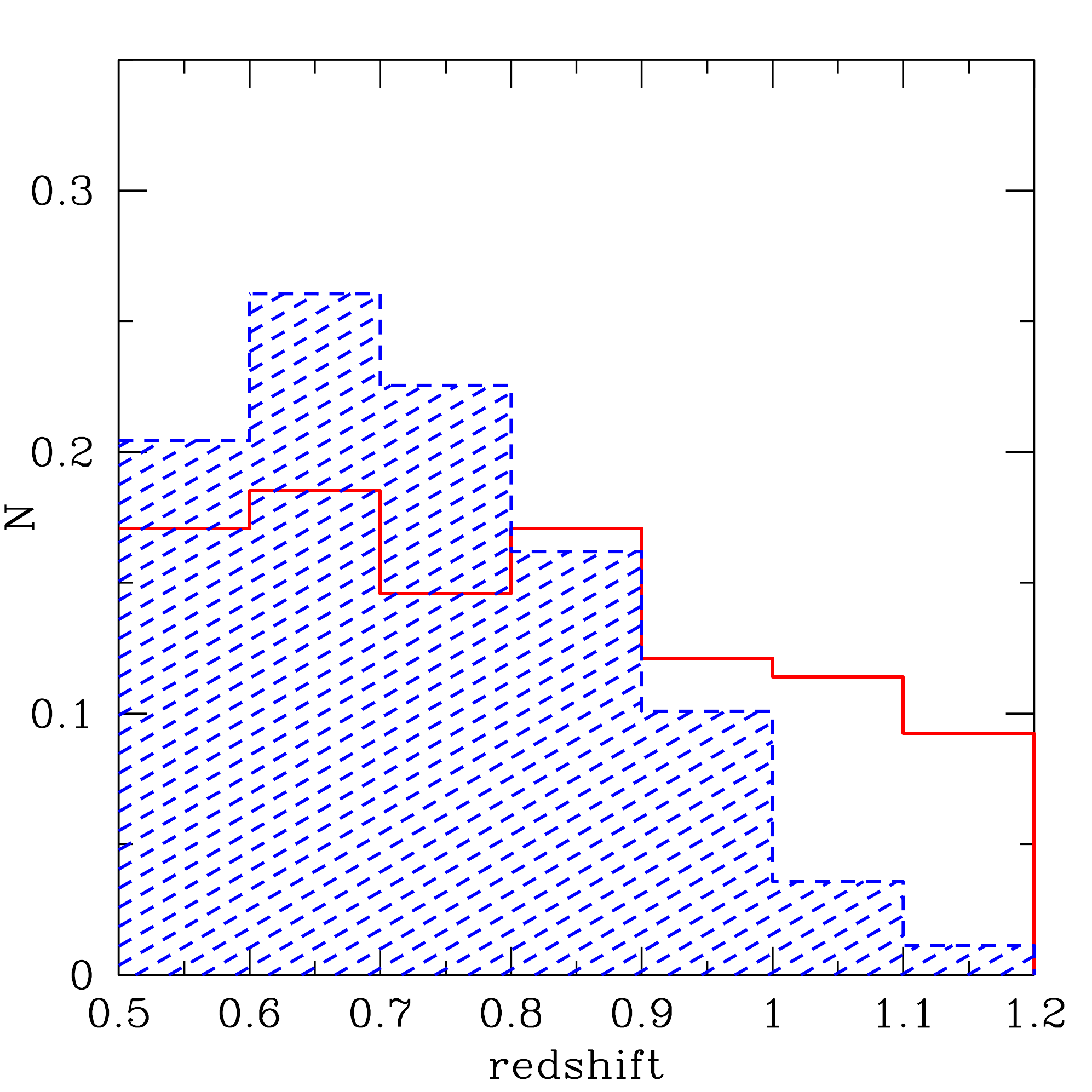}
\end{subfigure}
\caption{From left to right: stellar mass, star-formation rate and redshift normalized distributions of the 281 AGN (red line) and the 19,541 galaxies (blue shaded area), with CFHTLS+WISE photometry available. The AGN distribution peaks at lower M$_*$ and higher SFR and redshift values compared to the galaxy population.}
\label{fig_distrib}
\end{figure*}

\begin{figure}
\begin{center}
\includegraphics[height=1.\columnwidth]{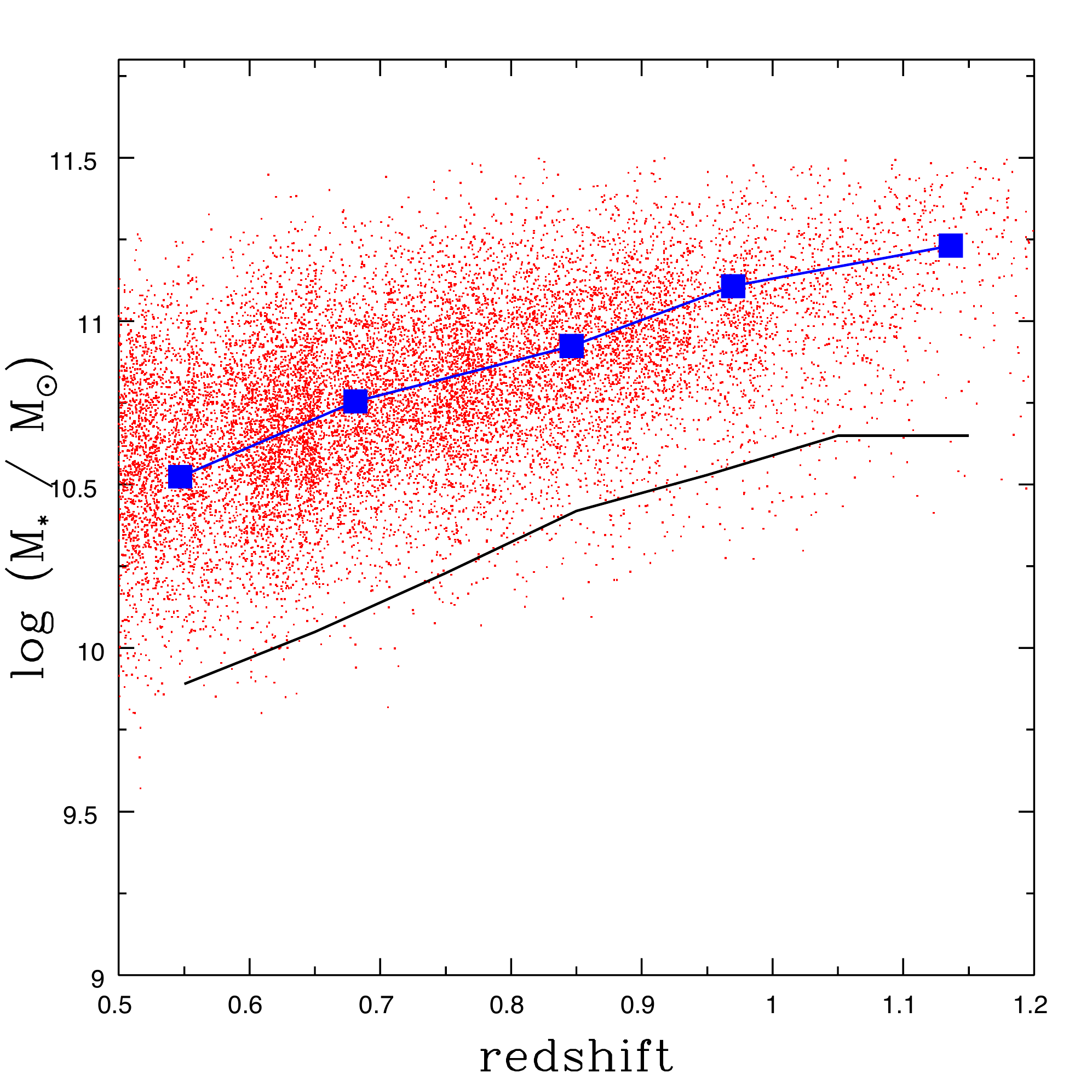}
\end{center}
\caption{The distribution of $M_{\star}$ as a function of redshift for the 19,541 galaxies of our sample. Individual sources are shown by red dots. The median stellar mass in bins of redshift is shown by the blue squares. The black line corresponds to the 95\% of the $M/L$ completeness limit (see text).}
\label{fig_mstar_redz}
\end{figure}

\begin{table*}
\caption{The X-ray AGN and galaxy samples in the VIPERS PDR-2 field within the redshift range of $0.5\leq z\leq 1.2$.}
\centering
\setlength{\tabcolsep}{1.5mm}
\begin{tabular}{lcccccc}
       \hline
 \hline
{AGN sample} & {galaxy sampe} & {$\rm <z>$} &{$\rm <\log L_X>$ }  & {b$_{CCF}$}  & {b$_{ACF}$} &{logM$_{DMH}$}\\
 & & & ($\rm erg \,s^{-1}$) & & & (h$^{-1}$\,M$_\odot$)  \\
       \hline
    \\
specz AGN  & 407 & 0.80 & 43.7  & {$1.54^{+0.08}_{-0.09}$}   &{$1.64^{+0.17}_{-0.20}$} & {$12.75^{+0.18}_{-0.26}$} \\
\\
galaxies &45,180  & 0.71 & & & {$1.39^{+0.04}_{-0.04}$}  & {$12.59^{+0.07}_{-0.05}$}  \\
		\\
       \hline
\label{table:clustering_values}
\end{tabular}
\end{table*}

\section{Data}

\subsection{Galaxy sample}
\label{galaxy_samples}
The galaxy population we use in our clustering analysis comes from the Public Data Release 2 \citep[PDR-2;][]{Scodeggio2016} of the VIPERS survey \citep{Guzzo2014, Garilli2014}. The observations have been carried out using the VIMOS \citep[VIsible MultiObject Spectrograph,][]{LeFevre2003} on the ESO Very Large Telescope (VLT). The survey covers an area of $\approx$ 23.5\,deg$^2$, split over two regions within the CFHTLS-Wide (Canada-France- Hawaii Telescope Legacy Survey) W1 and W4 fields. Follow-up spectroscopic targets were selected to the magnitude limit i$^\prime=22.5$  from the T0006 data release of the CFHTLS catalogues. An optical colour-colour pre-selection, i.e., {\it{[(r-i)$ > $0.5(u-g) or (r-i)$ > $0.7]}}, excludes galaxies at $z<0.5$, yielding a $>98\%$ completeness for $z>0.6$ \citep[for more details see][]{Guzzo2014}. PDR-2 consists of 86,775 galaxies with available spectra. Each spectrum is assigned a quality flag that quantifies the redshift reliability. In all VIPERS papers, redshifts with flags in the range between 2 and 9 are considered as reliable and are those used in the science analysis \citep{Garilli2014, Scodeggio2016}. Restricting the redshift range to 0.5$<$z$<$1.2 our final  sample consists of 45,180 galaxies.

To compare our AGN clustering dependence with that of galaxies with similar properties, we estimate SFR, $M_{\star}$ and sSFR for our galaxy sample using the CIGALE code version 0.12 \citep[Code Investigating GALaxy Emission;][]{Noll2009} to fit their Spectral Energy Distributions (SEDs; see Table \ref{table_cigale} and Section \ref{agn_samples} for more details on the SED fitting analysis). We construct galaxy SEDs using optical photometry from CFHTLS and mid-IR photometry from WISE. After applying our SED criteria (see Section \ref{agn_samples}) there are 42,412 galaxies with optical photometry. 19,541 (46\%) of them have additionally mid-IR photometry and 16,643 (39\%) sources have optical+near-IR (NIR) photometric bands (see Appendix \ref{append_nir}). Figure \ref{fig_distrib} presents the M$_\star$, SFR and redshift distributions of the 19,541 galaxies (blue shaded area). Fig. \ref{fig_mstar_redz} presents the distribution of $M_{\star}$ as a function of redshift. The median stellar mass in bins of redshift is shown by the blue squares. The minimum $M_{\star}$ for which the galaxy sample is complete depends on the redshift and the stellar mass-to-light ratio, $M/L$. To account for this limit we follow the methodology presented in \cite{Pozzetti2010}. In brief, for each galaxy we estimate the $M_{lim}$, i.e., the mass the galaxy would have, at its redshift, if its apparent magnitude were equal to the limiting magnitude of the survey. We then use the $M_{lim}$ of the 20\% of the faintest galaxies at each redshift and we derive the $M_{min}$, that is the minimum mass above which the derived galaxy stellar mass function is complete. The $M_{min}$ is defined as the upper envelope of the $M_{lim}$ distribution below which lie 95\% of the $M_{lim}$ values at each redshift. The results are shown by the black line in Fig. \ref{fig_mstar_redz}. Usage of the AGN/galaxy cross-correlation function (see Section \ref{sec_analysis}) alleviates the effect of this mass incompleteness on our clustering measurements.

\subsection{AGN sample}
\label{agn_samples}
The X-ray AGN sample used in our analysis comes from the XMM-XXL survey \citep{Pierre2016} that covers an area of about 50\,deg$^2$ with an exposure time of about 10\,ks per XMM pointing. In the equatorial subregion of the field ($\sim$  25\,deg$^2$) that overlaps with the CFHTLS W1 field, 8,445 X-ray sources are detected \citep{Liu2016}. Reliable spectroscopy from SDSS-III/BOSS  \citep{Eisenstein2011, Smee2013,Dawson2013} is available for 2,512 AGN \citep{Menzel2016}. Spectroscopic redshifts are complemented from the VIPERS PDR-2 spectroscopic data \citep{Scodeggio2016}. 1,192 of these sources are within the VIPERS PDR-2 mask (see Section \ref{galaxy_samples}). Restricting the redshift range to 0.5$<$z$<$1.2 to match that of our galaxy sample, results in 407 X-ray AGN, with $\log \, L_X (\rm 2-10\,\rm{keV}) \ge 42$\,  erg s$^{-1}$. 

Stellar masses and Star Formation Rates for our X-ray AGN sample, are estimated using the CIGALE code. SEDs are constructed using multiwavelength broadband photometric data, from optical (CFHTLS) to mid-infrared (WISE) wavelengths. The association of X-ray sources with optical and IR counterparts was based on the likelihood ratio method \citep[e.g.][]{Sutherland_and_Saunders1992}. The reduction of the {\it{XMM}} observations and the optical and IR identifications of the X-ray sources are described in detail in \cite{Georgakakis2017} and \cite{Menzel2016}. Adopting CFHTLS optical photometry instead of SDSS, decreases our X-ray sample by $\sim 10\%$, i.e., 364 sources compared to 407. However, for consistency with the galaxy sample (see Section \ref{galaxy_samples}) we use CFHTLS photometry in our SED analysis (see Appendix \ref{append_SDSS_CFHTLS}). Their X-ray luminosity distribution is presented in Fig. \ref{fig_lx}. Due to the large area of the XMM-XXL and the flux limit of the survey \citep[F$_{0.5-10\,\rm{keV}}>10^{-15}$\,erg\,cm$^{-2}$\,s$^{-1}$;][]{Menzel2016} our X-ray sample spans about an order of magnitude higher luminosities compared to smaller and deeper surveys, e.g. CDFS, AEGIS \citep[see e.g. Fig. 1 in][]{Fotopoulou2016}. Lack of far-infrared photometry does not affect our stellar mass estimations, but our SFR measurements are systematically underestimated \citep[see][]{Masoura2018}. This correction is neglected in our SFR measurements since it does not affect our clustering dependence analysis. Obscured AGN also contribute in the far-IR part of the SED \citep[10-15\%, e.g.][]{Risaliti2002}. Ignoring this contribution results in an overestimation of the SFR estimations in the case of absorbed sources. However, in our analysis we are not interested in absolute SFR values. Therefore the effect of this contribution on our clustering measurements should not be significant.

For the SED fitting we assume the double-exponentially-decreasing (2$\tau$-dec) model to convolve star-formation histories. \cite{Ciesla2015}  found that the 2$\tau$-dec model provides the best estimates of $M_{\star}$ and SFR of the simulated galaxies, in the expense of unrealistic galaxy ages. Using a different model, e.g. Single Stellar Population, does not affect our measurements. The AGN emission is modelled using the \cite{Fritz2006} library of template SEDs \citep{Ciesla2015}. We use the \cite{Bruzual_Charlot2003} for the stellar population synthesis models and assume the Salpeter Initial Mass Function (IMF) and the \cite{Calzetti2000} dust extinction law. For the absorbed dust that is reemitted in the IR we use the \cite{Dale2014} templates. Table \ref{table_cigale} presents the models and the values for their free parameters used by CIGALE for the SED fitting of the X-ray sample. Adopting different models, results in $\sim 0.1$\,dex variance in the estimated values of SFR, M$_\star$ and sSFR. The effect of different CIGALE libraries on our calculations is studied in detail in Appendix \ref{append_SED_libraries}.

To use only those AGN (and galaxies) with the most robust SED fitting measurements we restrict our sources to those with SED fits with $\chi ^2 _{red}<5$. There are 305 X-ray sources with available CFHTLS photometry that satisfy these criteria. The $\chi ^2 _{red}<5$ criterion excludes 38 X-ray sources from our AGN sample and has been chosen after visual inspection of the SED fits. Increasing the threshold to $\chi ^2 _{red}<6$, adds 10 more AGN to our dataset, but the vast majority of them have SEDs with bad fits and therefore unreliable M$_\star$ and SFR estimations. Reducing it to $\chi ^2 _{red}<4$ would exclude 15 more sources, most of them with reliable fits. Using additionally NIR photometry reduces our 305 sources to 239 {(78\%) and increases systematically the M$_{\star}$ calculations by $< 0.25$\,dex (both for the X-ray and the galaxy samples, see Appendix \ref{append_nir}). To obtain the most statistically robust correlation function measurements, we use the 305 AGN (only optical photometry) when studying the stellar mass clustering dependence. However, when our clustering analysis includes SFR calculations (e.g. SFR, sSFR) as well as when we compare the AGN and galaxy clustering (Section \ref{sec_galaxy_matched}) we require availability of mid-IR photometry. This, reduces our X-ray sample to 281 (92\%) sources. Figure \ref{fig_distrib} presents the stellar mass, SFR and redshift distributions of the 281 X-ray sources, respectively (solid lines). The AGN distribution (red line) peaks at lower M$_*$  and higher SFR and redshift values compared to the galaxy population (blue shaded area). These differences do not affect our clustering measurements since the galaxy contribution is "subtracted" from the cross-correlation signal to derive the DMH mass estimations (see Section 3). Moreover, when comparing the AGN and galaxy clustering (Section 4.6) we match the SFR, M$_*$ and redshift of the two samples.

\begin{table}
\caption{The properties of our X-ray AGN subsamples above and below the galaxy main sequence (see Section \ref{sec_main_sequence} for more details).}
\centering
\setlength{\tabcolsep}{0.3mm}
\begin{tabular}{ccc}
       \hline
$\log$ SFR$_{\rm{norm}}$ & $>0$ & $<0$  \\
	\hline
number of X-ray AGN & 123 & 158 \\
$\log$ SFR (M$_{\odot}\, \rm{yr}^{-1})$ & 1.38 & 0.95 \\
$\log (M_{\star} / M_{\odot})$ &10.1 & 10.7  \\
sSFR (\rm{G\rm{yr}}$^{-1})$  & 0.24 & -0.75 \\
$\log \, L_X (\rm 2-10\,\rm{keV})$ & 43.8& 43.3   \\
redshift &0.87 &  0.76   \\
$\log M / (M_{\odot} \,  h^{-1})$ & $12.90^{+0.32}_{-0.30}$ & $12.81^{+0.25}_{-0.22}$   \\
       \hline
\label{table:sfr_norm}
\end{tabular}
\end{table}

\section{Clustering analysis}
\label{sec_analysis}
The clustering analysis we follow is the same described in \cite{Mountrichas2016}. In summary, the projected correlation function is calculated, $w_p(\sigma)$:

\begin{equation}
w_p(\sigma)=2\int_0^\infty \xi(\sigma,\pi)d\pi, 
\label{eqn:wp}
\end{equation}
\noindent where $\sigma$ is the distance between two objects perpendicular to the line of sight and $\pi$ the separation along the line of sight. The upper limit of the integral is determined by computing $w_p(\sigma)$ for $\pi_{max}$ in the range 10-100\,Mpc. The clustering signal saturates for $\pi_{max}=20$\,Mpc and $\pi_{max}=50$\,Mpc for the AGN/galaxy cross-correlation and the galaxy autocorrelation functions, respectively. Smaller values lead to an underestimation of the clustering signal whereas larger values increase the noise from uncorrelated pairs. 

The calculated 2-halo term of the projected correlation function,  $w_{p}^{2h}(\sigma)$,  is connected with the projected correlation function of dark matter, $w_{DM}^{2h}(\sigma)$, via the following equation:

\begin{equation}
w_{p}^{2h}(\sigma)=b^2w_{DM}^{2h}(\sigma).
\label{eqn:proj_dm}
\end{equation}
\noindent In the case of a galaxy autocorrelation function, $b=b_g$, i.e., the galaxy bias, whereas in the case of an AGN/galaxy cross-correlation, $b=b_{AG}$, i.e., the AGN/galaxy bias. The best-fit bias in the above equation is defined by applying a $\chi^2$ minimization $\chi^2=\Delta^{T}M_{cov}^{-1}\Delta$, on scales 1.5-20\,h$^{-1}$Mpc, where $M_{cov}^{-1}$ is the inverse of the covariance matrix, that quantifies the degree of correlation between the different bins of $w_p(\sigma)$. $\Delta$ is defined as $w_{p,2h}-w_{p,model}$, where $w_{p,model}=b^2w_{DM}^{2h}(\sigma)$. The AGN bias can then be inferred via

\begin{equation}
b_{AGN}=\frac{b_{AG}^2}{b_g}.
\end{equation}

The Dark Matter Halo Mass (DMHM) is calculated by adopting the ellipsoidal  model of   \cite{Sheth2001} and the analytical approximations of \cite{Bosch2002}.

The inferred DMHM depends on the scales of the correlation function that are fitted. In our case there is a $\sim 10\%$ fluctuation in the DMHM estimations when the correlation function is fitted from 1.5-10\,h$^{-1}$Mpc, instead of the 1.5-20\,h$^{-1}$Mpc used in our analysis. An additional source of uncertainty comes from the size of our X-ray sample. The AGN dataset is not large enough to allow us model the contribution of sources from the same DMH (1-halo term of the correlation function). Thus, we only model the 2-halo term. This is further discussed in Section \ref{sec_discussion}. Finally, although XXM-XXL covers $\sim 25$\,deg$^2$, our measurements are still affected by cosmic variance. In Section \ref{sec_discussion} and Appendix \ref{append_vipers} we discuss in more details the effect of cosmic variance on our estimations.

The uncertainties of our auto-correlation and cross-correlation measurements are calculated using the Jackknife methodology \citep[e.g.][]{Mountrichas2016}. The area of the survey is split into $N_{JK}=40$ sections  and  the correlation function  is measured  $N_{JK}$ times  by excluding  one  section each time. All the errors presented in the rest of this work are based on the Jackknife method, unless otherwise stated.

For the estimation of the correlation functions, we generate a random catalogue that has the same sky footprint with the galaxy sample. To achieve that we pass it through  both the photometric mask (i.e. bright stars) and the spectroscopic mask (Field-of-View of the spectrograph). Our random sample is $\approx 20\times$ the number of galaxies. This is a large enough number to minimise the shot noise contribution to the estimated correlation functions while at the same time the catalogue's size allows computation efficiency \citep{Mountrichas2016}.

\begin{table*}
\caption{Relative bias estimations from our cross-correlation function measurements for the different X-ray AGN subsamples. For the definition of the relative bias in each case, see the relevant Sections.}
\centering
\setlength{\tabcolsep}{1.mm}
\begin{tabular}{ccccccc}
       \hline
 & SFR$_{\rm{norm}}$ & $\log (M_{\star} / M_{\odot})$ & $\log (M_{\star} / M_{\odot})$ & $\log$\,SFR (M$_{\odot}\, \rm{yr}^{-1}$) & $\log$\,sSFR (\rm{G\rm{yr}}$^{-1})$  & AGN/galaxies (matched)  \\
& 0.0 & 10.5 & 10.8 & 1.1  & -0.4  &  \\

	\hline
$b_{rel}$ &$0.97^{+0.12}_{-0.11}$& $1.28^{+0.15}_{-0.14}$ & $1.20^{+0.10}_{-0.10}$ &$1.23^{+0.13}_{-0.14}$ & $1.31^{+0.16}_{-0.16}$&  $1.03^{+0.07}_{-0.06}$\\

	\hline
\label{table:relative bias}
\end{tabular}
\end{table*}

\begin{figure}
\begin{center}
\includegraphics[height=1.\columnwidth]{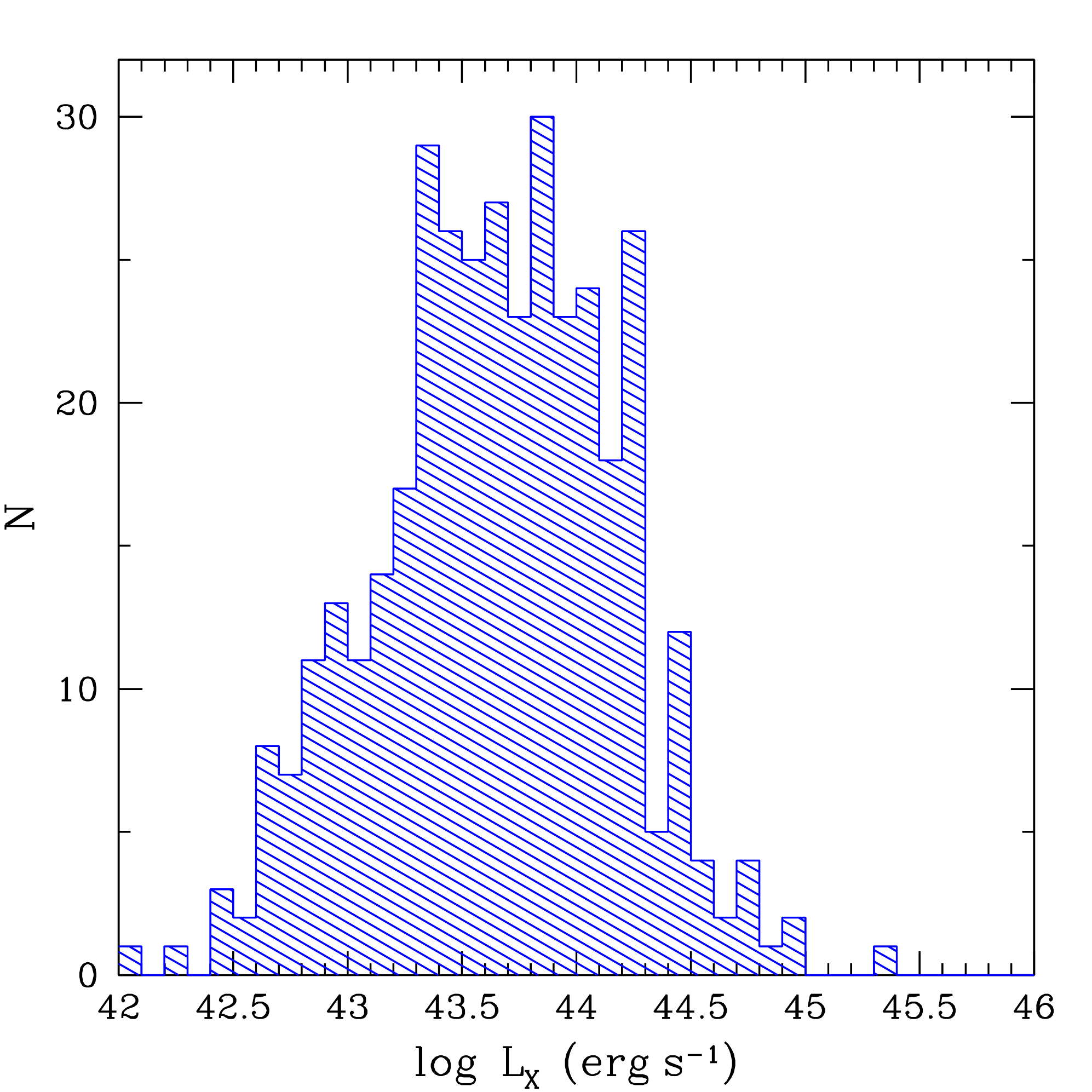}
\end{center}
\caption{The X-ray luminosity distribution of the 364 X-ray selected AGN sample, with CFHTLS photometry available, in the redshift interval $0.5<z<1.2$. The median luminosity of the sample is $\log L_X (\rm  2-10\,\rm{keV})=43.7\, erg \,s^{-1}$.}
\label{fig_lx}
\end{figure}

\section{Results}
\label{results}

First, we measure the DMH mass of the total sample of X-ray AGN within the VIPERS PDR-2 region. The goal is to compare the measurements presented in the current work with those in \cite{Mountrichas2016} in the PDR-1 region. The two studies use the same X-ray data but the area that overlaps with the galaxy sample has been doubled in this work. Next, we study the dependence of the X-ray clustering on various properties of the host galaxy, i.e., M$_{\star}$, SFR and sSFR. In each of these cases, the purpose is to study the dependence of AGN clustering on one parameter and at the same time avoid the degeneracy among the galaxy properties. To accomplish this, the various cuts applied are based on two criteria: (i) create nearly equal subsamples and (ii) only one of the host galaxy properties differs between two subsamples while the rest of the properties have similar median values. This allows us to attribute any clustering difference to the galaxy property under consideration. We also measure the DMH mass of X-ray AGN as a function of the location of the host galaxy in respect to the star-forming main sequence. Finally, we create matched AGN and galaxy samples and compare their clustering properties.

\subsection{The AGN/galaxy clustering in the VIPERS field}

We calculate the correlation function and derive the DMH mass of the full X-ray AGN and galaxy samples, i.e. no M$_{\star}$ restriction or SED fitting criterion are applied (Table \ref{table:clustering_values}). 

Fig. \ref{clustering_measurements} presents the AGN/galaxy cross-correlation (left panel) and galaxy autocorrelation (right panel) measurements, for the total AGN sample of 407 X-ray sources. For the determination of the correlation functions we have taken into account the spectroscopic window function and the sampling rate of the VIPERS galaxy sample \citep{Scodeggio2016} following the methods described in \cite{delaTorre2017} and \cite{Pezzotta2017}. The fraction of sources with a measured spectrum is defined as the Target Sampling Rate (TSR) while the fraction of observed spectra with reliable redshift measurement as the Spectroscopic Sampling Rate (SSR). These two factors correct for incompleteness at large scales when measuring either the galaxy autocorrelation function or the AGN/galaxy cross-correlation function. In addition, the MOS slit length imposes a minimum angular separation in the selection of spectroscopic targets. Although this effect only affects measurements below $\approx$ 1\,h$^{-1}$Mpc, i.e. smaller than those probed in our analysis (1.5-20\,h$^{-1}$Mpc), we take it into account in the correlation function measurements, by comparing the angular auto-correlation function of the target and spectroscopic samples. The methodology used in this study to account for both of these effects is identical to that presented in \cite{Mountrichas2016} and is explained in more detail in their Section 3.2. The only difference is that in the PDR-2 the estimation of SSR and TSR has been improved. Specifically the SSR, now accounts for new galaxy property dependencies \citep{Scodeggio2016}, while the TSR has been recomputed with better angular resolution \citep{delaTorre2017}.  

\cite{Mountrichas2016} used the PDR-1 of VIPERS and performed a similar clustering analysis. PDR-1 covers approximately half the area of PDR-2. 20,109 galaxies were used to estimate their cross-correlation with 318 X-ray AGN, in the same redshift range with that used in this study. Their results are shown in Fig. \ref{clustering_measurements} (blue symbols). Their correlation function estimations appear lower compared to the measurements of this work (red symbols). Further investigation reveals that the reasons of this difference are the larger area covered by the PDR2 sample and the improved estimations of the TSR and SSR parameters. This is further discussed in Section \ref{sec_discussion} and Appendix \ref{append_vipers}.

\subsection{Comparison of the X-ray clustering above and below the galaxy main sequence}
\label{sec_main_sequence}

The correlation between SFR and M$_*$ is known as the main sequence (MS) of star-forming galaxies \citep[e.g.][]{Noeske2007}. Based on their location in the MS plane, galaxies are identified as star-forming (above the MS) and quiescent (below the MS). Previous studies have found different clustering properties for galaxies that lie below and above the MS \citep[e.g.][]{Coil2017}. Recent results indicate that the nearly linear slope of the SFR-M$_*$ relation becomes flatter at high masses ($\log (M_{\star} / M_{\odot})>$10.5), at least at $z<2$ \citep[see Fig. 10 in ][]{Schreiber2015}. These results, along with the intrinsic scatter of the MS, caused by e.g. the selection of the various Star-Formation History (SFH) and dust recipes, may misplace an AGN in relation to the MS. Therefore, to compare the  X-ray clustering above and below the MS we disentangle the effects of $M_{\star}$ and redshift on SFR. For that purpose, we divide the SFR of X-ray sources with the SFR of MS galaxies with the same stellar mass and redshift, using expression 9 in \cite{Schreiber2015}. We call this ratio normalized SFR, SFR$_{\rm{norm}}$ \citep[see][]{Masoura2018}. Therefore, the SFR$_{\rm{norm}}$ parameter indicates whether an AGN host galaxy with a specific M$_*$ at a specific redshift has higher (SFR$_{\rm{norm}}>$0) or lower (SFR$_{\rm{norm}}<$0) SFR than a normal (non-active) galaxy with the same M$_*$ and redshift.

Fig. \ref{fig_sfr_norm} presents the measurements for the two AGN subsamples. The mean properties of the two populations are shown in Table \ref{table:sfr_norm}. The relative bias, defined as the square root of the ratio of the AGN above the galaxy main sequence (SFR$_{\rm{norm}}>0$) to the AGN below the main sequence (SFR$_{\rm{norm}}<0$), is shown in Table \ref{table:relative bias} ($b_{rel}=\sqrt{{w_p(\sigma)_1}/{w_p(\sigma)_2}}$; e.g. Mountrichas \& Georgakakis 2012). Our results show that there is no statistical difference in the clustering of X-ray AGN that lie above and below the main sequence.

\subsection{X-ray clustering dependence on M$_\star$}

In this Section, we study the X-ray clustering dependence on stellar mass. For that purpose we split our 305 X-ray AGN into high and low $M_{\star}$ subsamples using two cuts, i.e., $\log (M_{\star} / M_{\odot})=10.5, 10.8$. These cuts have been chosen because they create  subsamples with very similar properties (SFR, redshift and X-ray luminosity; see Table \ref{table:stellar_mass}).

Fig. \ref{fig_clus_dependence} (left panel) presents the AGN/galaxy cross-correlation functions applying a $\log (M_{\star} / M_{\odot})=10.5$ cut. The high stellar mass subsample (circles) has a stronger clustering signal compared to the low stellar mass subsample. This is also the case when applying a cut at $\log (M_{\star} / M_{\odot})=10.8$, as shown by the DMHM (Table \ref{table:stellar_mass}) and relative bias estimations (Table \ref{table:relative bias}). The relative bias, b$_{rel}$, is defined as the square root of the ratio of the correlation functions of high to low stellar mass subsamples.

\subsection{X-ray clustering dependence on SFR}

Next, we study the dependence of the X-ray AGN clustering on the star-formation rate of their host galaxies. As mentioned in the previous Section, for these measurements we use the X-ray sample with the 281 sources that have optical and mid-IR photometry available. We apply a $\log$ SFR (M$_{\odot}\, \rm{yr}^{-1})=1.1$ to divide the AGN sources into two nearly equal subsamples and estimate their corresponding cross-correlation function with the VIPERS galaxies. The results are presented in Tables \ref{table:relative bias}, \ref{table:sfr} and Fig. \ref{fig_clus_dependence} (middle panel). The relative bias in this case, is defined as the square root of the ratio of the low to high SFR. The two SFR subsamples have a 0.5 dex difference in X-ray luminosity that could also contribute to the difference in the DMH mass estimations. For that purpose we apply a narrower redshift and X-ray luminosity range on the 281 X-ray AGN, i.e., $0.6 \leq z \leq 1.1$ and $43 \leq \log L_{X (2 -10\,\rm{keV}}) \leq 44.2$, restricting the sample to 159 sources. This particular cut eliminates the L$_X$ difference and excludes the smallest possible number of sources from the X-ray sample. Although the sample is limited and the estimations have large uncertainties (Table \ref{table:sfr}), the trend we observe is consistent with the indications from the previous measurements, i.e., a negative X-ray AGN clustering dependence on SFR.

\subsection{X-ray clustering dependence on sSFR}

In this Section we study the dependence of X-ray AGN clustering on specific SFR. Fig. \ref{fig_clus_dependence} (right panel) presents the correlation function measurements when splitting the 281 X-ray AGN sample into low and high sSFR subsamples, using a $\log$ sSFR$ (\rm{G\rm{yr}}^{-1})=-0.4$ cut. This cut correspond to $\log (M_{\star} / M_{\odot})=10.5$ and  $\log$ SFR (M$_{\odot}\, \rm{yr}^{-1})=1.1$ cuts used in the previous Sections and also splits the X-ray dataset into two nearly equal subsets. The properties of the two subsamples and the inferred DMH masses are shown in Table \ref{table:ssfr}. Table \ref{table:relative bias} presents the relative bias estimation, defined as the square root of the ratio of the correlation function of AGN with $\log$ sSFR$ (\rm{G\rm{yr}}^{-1})<-0.4$ over the correlation function of AGN with $\log$ sSFR$ (\rm{G\rm{yr}}^{-1})>-0.4$. A negative dependence of the X-ray clustering on sSFR is detected. 

\begin{figure*}
\centering
\begin{subfigure}
  \centering
  \includegraphics[width=.49\linewidth]{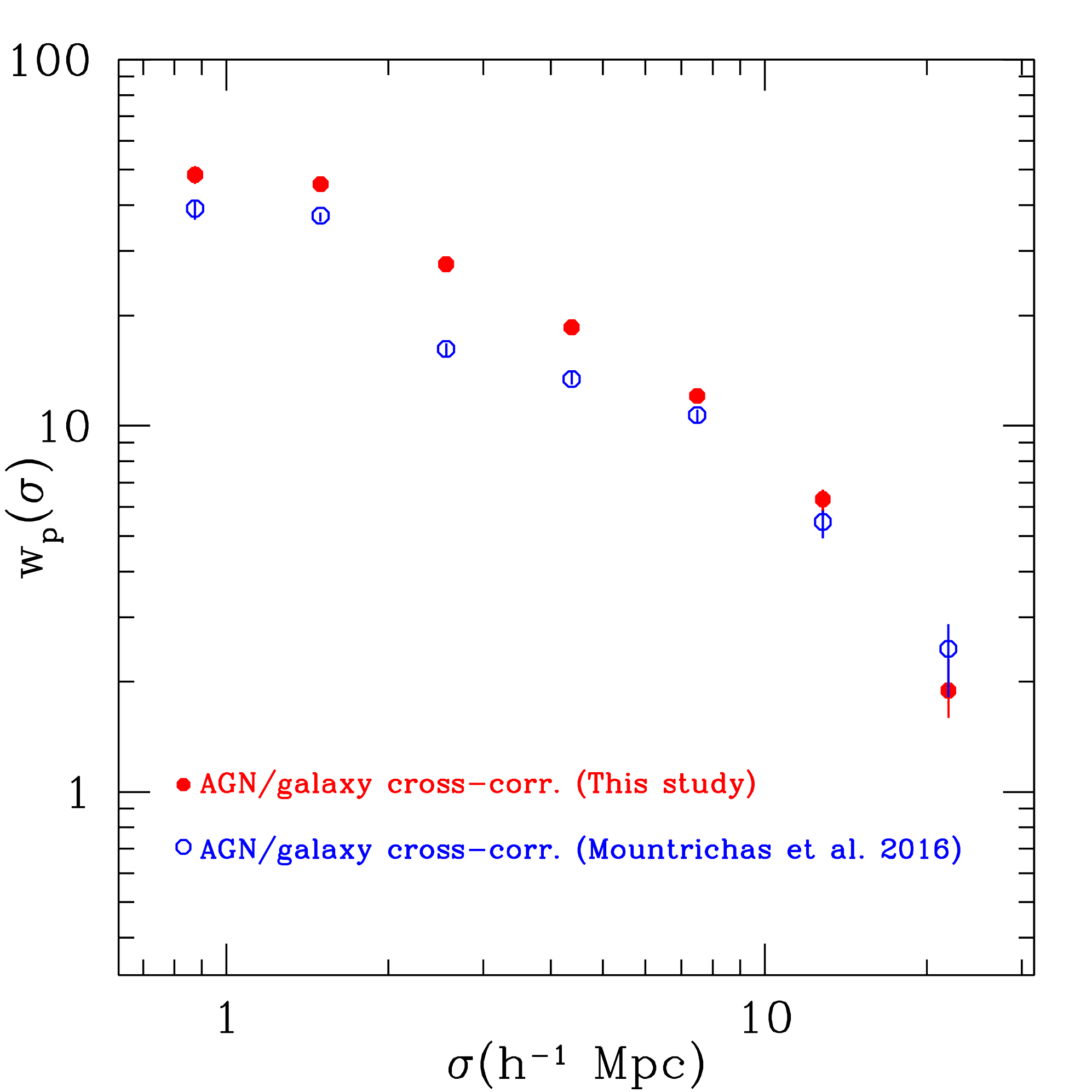}
\end{subfigure}
\begin{subfigure}
  \centering
  \includegraphics[width=.49\linewidth]{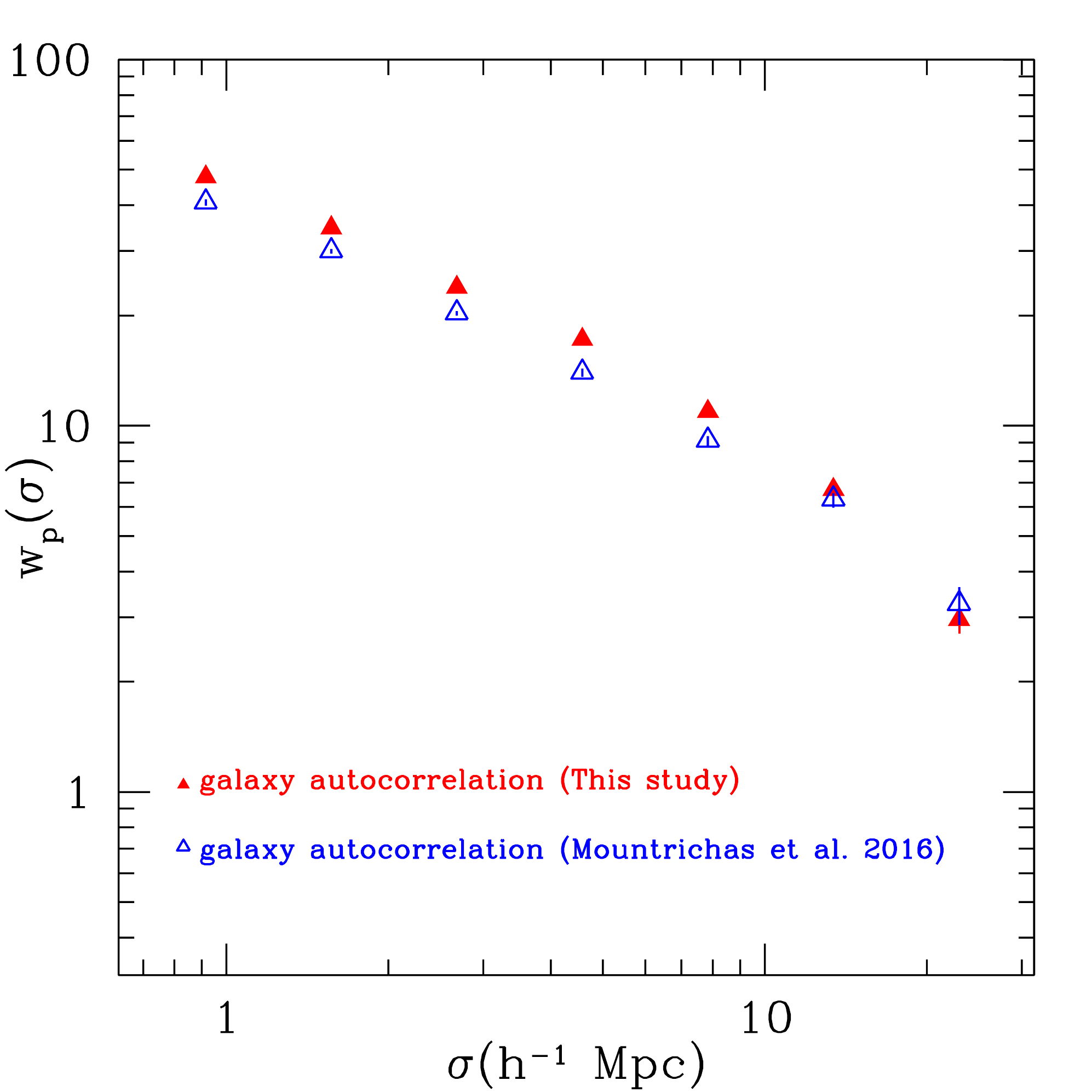}
\end{subfigure}
\caption{Left: The AGN/galaxy cross correlation function measurement using the PDR-2 of the VIPERS galaxy catalogue (red circled). Blue circles present the measurements from Mountrichas et al. 2016, using the PDR1 of the VIPERS survey. Right: The galaxy autocorrelation function estimation with the PDR2 galaxy sample (red triangles) compared to the Mountrichas et al. 2016 measurements (PDR-1; blue triangles). The PDR-2 measurements appear slightly higher compared those using the PDR-1. This could be due to the cosmic variance, since PDR-2 covers approximately double the area covered by PDR-1. The inferred bias and DMHM estimations are in statistical agreement.}
\label{clustering_measurements}
\end{figure*}

\subsection{Comparison of the X-ray AGN and galaxy clustering}
\label{sec_galaxy_matched}

In this Section, we compare the clustering of X-ray selected AGN with that of normal galaxies. The goal is to test whether the presence of an active supermassive black hole affects the large-scale structure of the host galaxy. 

We use the 281 X-ray AGN and 19,541 galaxies with available CFHTLS+WISE photometry and we construct galaxy samples that match the $M_{\star}$, SFR and redshift distribution of the X-ray dataset (Fig. \ref{fig_distrib}). For that purpose, we join the three distributions, using bins of $\Delta \rm{log}(M_{\star}/ M_{\odot})=0.25$, $\Delta \rm{log}SFR (M_{\odot}\, \rm{yr}^{-1})=0.25$ and $\Delta z=0.1$ and we normalize each one by the total number of sources in each bin. This effectively gives us the PDF (probability density function) in this 3-D parameter space. Then we use the M$_\star$, SFR and redshift of each galaxy to weigh it based on this estimated PDF \citep{Mendez2016}. 

Fig. \ref{fig_matched} presents the correlation function measurements. The estimated DMH masses are, $\log M / (M_{\odot} \,  h^{-1})=12.85^{+0.28}_{-0.25}$ and  $\log M / (M_{\odot} \,  h^{-1})=12.83^{+0.09}_{-0.09}$, for the AGN and galaxy populations, respectively. The results show that the AGN clustering is similar with the clustering of normal galaxies that have the same  M$_\star$, SFR and redshift distributions with the X-ray sources. 

Then, we repeat the analysis we followed for the X-ray sources, i.e., we split the (matched) galaxy sample using  M$_\star$, SFR and sSFR cuts. The results are presented in Table \ref{table:matched_gals}. In Fig. \ref{rel_bias_all} we plot the variation of the relative bias as a function of separation for each galaxy property, i.e., M$_\star$ (top panel), SFR (middle panel) and sSFR (bottom panel). We find that the clustering of normal galaxies (blue shaded regions) have similar dependence on galaxy properties with the clustering of X-ray AGN (green shaded regions).


\begin{figure}
\begin{center}
\includegraphics[height=1.\columnwidth]{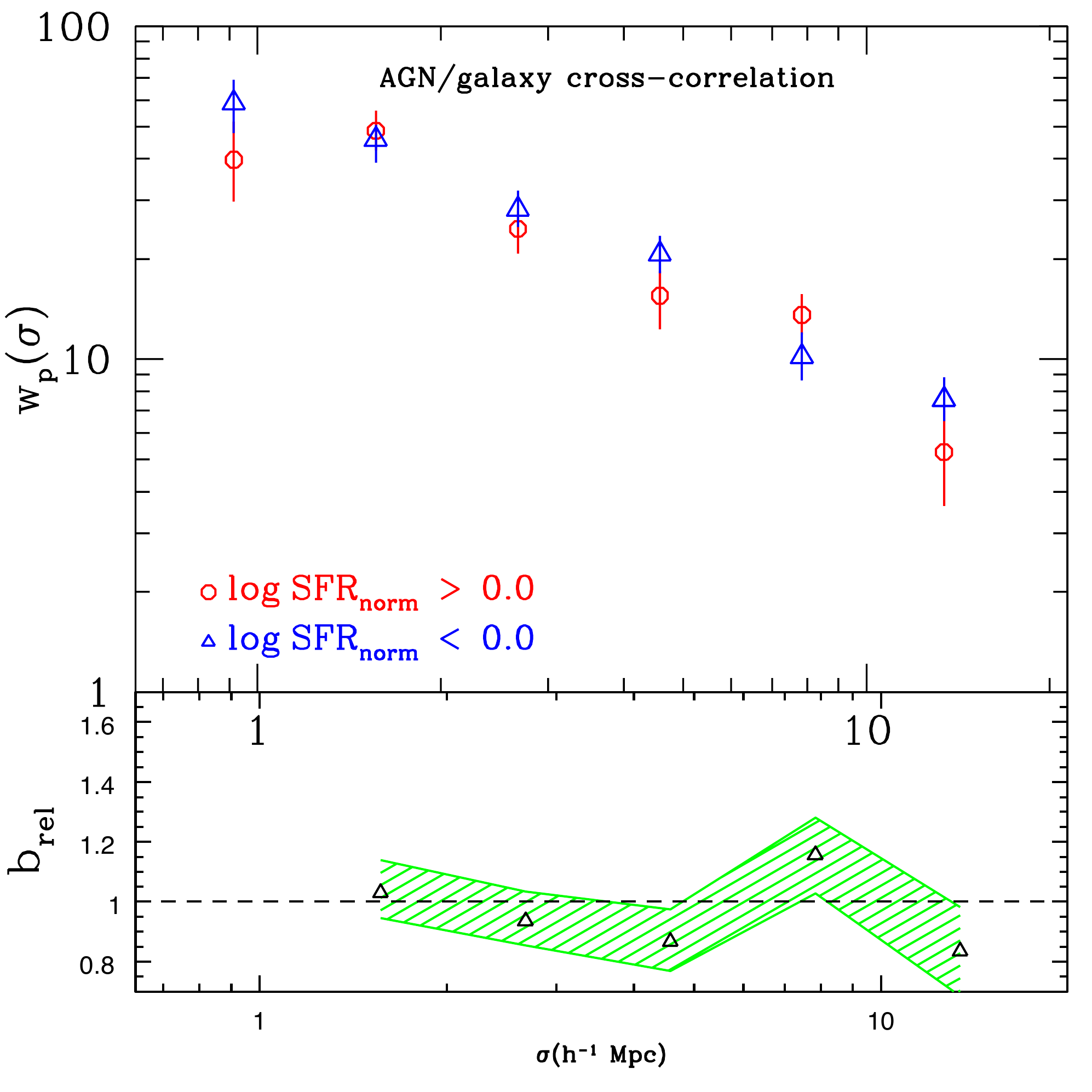}
\end{center}
\caption{Top panel: The AGN/galaxy cross-correlation functions for AGN with $\log$ SFR$_{norm}>0.0$ (open circles) and $\log$ SFR$_{norm}<0.0$ (open triangles). X-ray AGN hosted by galaxies that lie above the galaxy main sequence of star-formation have similar clustering with AGN that live in galaxies that are below the galaxy main sequence. Bottom panel: The relative bias calculations as a function of distance.}
\label{fig_sfr_norm}
\end{figure}

\begin{table}
\caption{The properties of the X-ray AGN subsamples, using different stellar mass cuts.}
\centering
\setlength{\tabcolsep}{0.5mm}
\begin{tabular}{ccccc}
       \hline
$\log (M_{\star} / M_{\odot})$ & $>10.5$ & $<10.5$ & $>10.8$ & $<10.8$  \\
mean $\log (M_{\star} / M_{\odot})$ & [10.8] & [10.1] & [11.0] & [10.3]  \\
	\hline
number of X-ray AGN & 160 & 145 & 72 & 233  \\
$\log$ SFR (M$_{\odot}\, \rm{yr}^{-1})$ &1.12 & 1.11 & 1.14 & 1.11 \\
$\log \, L_X (\rm 2-10\,\rm{keV})$ & 43.6& 43.4 & 43.6 & 43.5  \\
redshift  &  0.79 & 0.82 & 0.83 & 0.80 \\
$\log M / (M_{\odot} \,  h^{-1})$ & $13.05^{+0.28}_{-0.31}$ & $12.67^{+0.22}_{-0.29}$ &$12.97^{+0.42}_{-0.67}$& $12.79^{+0.25}_{-0.24}$  \\
       \hline
\label{table:stellar_mass}
\end{tabular}
\end{table}

\begin{table}
\caption{The properties of the X-ray AGN subsamples when we split them based on their Star Formation Rate. The SFR cut marked with an asterisk indicates the measurements when we restrict the AGN sample to $0.6 \leq z \leq 1.1$ and $43 \leq {\rm log} L_X (2 - 10\,\rm{\rm{keV}}) \leq 44.2$ (see Section 4 for more details).}
\centering
\setlength{\tabcolsep}{0.5mm}
\begin{tabular}{ccccc}
       \hline
$\log$ SFR (M$_{\odot}\, \rm{yr}^{-1})$ & $>1.1$ & $<1.1$ & $>1.1*$ & $<1.1*$ \\
mean $\log$ SFR (M$_{\odot}\, \rm{yr}^{-1})$ & [1.41] & [0.79 ] & [1.41] & [0.78 ]\\
	\hline
number of X-ray AGN & 158 & 123 & 89 & 70  \\
$\log (M_{\star} / M_{\odot})$ &10.5 & 10.5 &10.5 & 10.4 \\
$\log \, L_X (\rm 2-10\,\rm{keV})$ & 43.7& 43.2 & 43.7& 43.7  \\
redshift &0.88 &  0.71 &0.87 &  0.77  \\
$\log M / (M_{\odot} \,  h^{-1})$ & $12.54^{+0.22}_{-0.18}$ & $13.08^{+0.38}_{-0.32}$ & $12.67^{+0.39}_{-0.75}$ & $13.16^{+0.57}_{-0.61}$  \\
       \hline
\label{table:sfr}
\end{tabular}
\end{table}

\begin{table}
\caption{The properties of the X-ray AGN subsamples when a $\log$ sSFR$=-0.4$\,(\rm{G\rm{yr}}$^{-1})$  cut is applied on the X-ray sample.}
\centering
\setlength{\tabcolsep}{0.3mm}
\begin{tabular}{ccc}
       \hline
$\log$ sSFR$ (\rm{G\rm{yr}}^{-1})$ & $>-0.4$ & $<-0.4$  \\
mean $\log \rm{sSFR} (\rm{G\rm{yr}}^{-1})$ & [0.17] & [-0.83] \\
	\hline
number of X-ray AGN & 145 & 136 \\
$\log$ SFR (M$_{\odot}\, \rm{yr}^{-1})$ & 1.35 & 0.91 \\
$\log (M_{\star} / M_{\odot})$ &10.2 & 10.8  \\
$\log \, L_X (\rm 2-10\,\rm{keV})$ & 43.7& 43.2   \\
redshift &0.87 &  0.74   \\
$\log M / (M_{\odot} \,  h^{-1})$ & $12.39^{+0.25}_{-0.21}$ & $13.18^{+0.40}_{-0.35}$   \\
       \hline
\label{table:ssfr}
\end{tabular}
\end{table}

\section{Discussion}
\label{sec_discussion}

In Fig.  \ref{clustering_measurements}, we compare the correlation function measurements using the VIPERS PDR-2 region (filled symbols) with the estimations from \cite{Mountrichas2016}, where they used the PDR-1 of the VIPERS catalogue (open symbols). Their DMH mass estimation (logM$_{DMH}=12.50^{+0.22}_{-0.30}$\,$h^{-1}$\,M$_\odot$) is about 0.5\,dex lower compared to estimates in the literature of the DMH mass of moderate luminosity AGN. They attribute this difference to the higher luminosity of their X-ray sample ($\rm
  log\,  L_X  (\rm   2-10\,\rm{keV})=  43.6^{+0.4}_{-0.4}\, erg\,s^{-1}$), i.e, they claim a negative dependence of the X-ray clustering on luminosity. Our DMH mass estimations (Table \ref{table:clustering_values}) are within the 1$\sigma$ uncertainties of the measurements. The galaxy autocorrelation function is higher when the PDR-2 dataset is utilized compared to the PDR-1 measurement. This also reflects in the AGN/galaxy cross-correlation measurements. The corresponding bias estimations, presented in Table \ref{table:clustering_values} appear $\sim$ 14\% higher in our study compared to the values presented in Table 1 of \cite{Mountrichas2016}. The difference may be attributed to the larger area ($\sim 2\times$) covered in PDR-2 compared to PDR-1 (cosmic variance) and/or to the improvement in the estimation of the SSR and TSR parameters. In Appendix \ref{append_vipers}, we explore in detail how these two factors contribute to the correlation function measurements and we show that the difference is mostly due to the increased area covered by the PDR-2 sample and to a lesser degree to the improved estimations of the TSR and SSR parameters. Specifically, using the updated values for the TSR and SSR increases the galaxy bias estimations by $\approx 3\%$, whereas due to cosmic variance the galaxy bias increases by  $\approx 8\%$.

Then, we studied the dependence of AGN clustering on various properties of the host galaxy. Our analysis revealed a dependence of the X-ray clustering on stellar mass and SFR ($\sim 0.4$\,dex difference in DMH mass) and a stronger dependence on sSFR ($\sim 0.8$\,dex difference in DMH mass). The X-ray luminosity is another parameter that the AGN clustering may depend on. Many previous clustering works have studied this dependence, with controversial results \citep[e.g.][]{Coil2009, krumpe2010, Krumpe2012, Krumpe2018, Mountrichas2012, Koutoulidis2013}. The X-ray sample is not large enough to improve the statistical significance of these previous studies and therefore we do not attempt to study the dependence of X-ray clustering on luminosity. However, as already mentioned, our DMH mass estimation agrees with that of \cite{Mountrichas2016} based on which they claim a negative dependence of AGN clustering with X-ray luminosity, at least within the X-ray luminosities spanned by our (and their) X-ray sample(s) ($\rm log\,  L_X  (\rm   2-10\,\rm{keV})\sim  43.6-43.7\,  erg\,s^{-1}$)

Moreover, within the error budget of the measurements, we find no significant difference in the environments between AGN that lie above and below the star-formation main sequence. This lack of clustering difference, could be attributed to the small size of the X-ray (sub-)samples (e.g., sample variance, statistical uncertainties). However, it could also suggest that additionally to the physical properties of the host galaxy, the DMH mass of X-ray AGN also depends on the location of the host in respect to the star-forming main sequence. \cite{Coil2017} found that, at a given stellar mass, star-forming galaxies above the main sequence with higher sSFR are more clustered than galaxies below the main sequence with lower sSFR.  For the quiescent galaxy population, their analysis revealed that sources with higher sSFR are less clustered than galaxies with lower sSFR. To test these scenarios we need to split the X-ray samples above and below the main sequence based on their host galaxy properties, i.e. $M_{\star}$, SFR and sSFR. However, the number of the X-ray AGN does not allow such an analysis.

\begin{table*}
\caption{DMH mass estimations for the matched galaxy samples. The clustering properties of normal galaxies show similar trends with those of X-ray AGN, i.e., a mild dependence on $M_{\star}$ and a strong dependence on sSFR.}
\centering
\setlength{\tabcolsep}{1mm}
\begin{tabular}{ccccccccc}
       \hline
$\log$ & $(M_{\star} / M_{\odot})$ & $(M_{\star} / M_{\odot})$  & $(M_{\star} / M_{\odot})$ & $(M_{\star} / M_{\odot})$  & SFR (M$_{\odot}\, \rm{yr}^{-1}$) & SFR (M$_{\odot}\, \rm{yr}^{-1}$) &  sSFR (\rm{G\rm{yr}}$^{-1})$ & sSFR (\rm{G\rm{yr}}$^{-1})$ \\
& $>10.5$ & $<10.5$ & $>10.8$ & $<10.8$ & $>1.1$ & $<1.1$ & $>-1.0$ & $<-1.0$ \\
	\hline
number of galaxies  & 10,550 & 8991 & 16,354 & 3,187 & 5,875 & 13,666 & 8,944 & 10,597 \\
$\log$ DMHM$_{\rm{gal}} / (M_{\odot} \,  h^{-1})$&$13.13^{+0.10}_{-0.12}$ & $12.77^{+0.06}_{-0.06} $ &$13.02^{+0.07}_{-0.09}$ & $12.81^{+0.13}_{-0.15}$  & $12.59^{+0.11}_{-0.11}$ & $13.04^{+0.10}_{-0.08}$ & $12.51^{+0.08}_{-0.09}$ & $13.24^{+0.13}_{-0.13}$  \\

       \hline
\label{table:matched_gals}
\end{tabular}
\end{table*}

\begin{figure*}
\centering
\begin{subfigure}
  \centering
  \includegraphics[width=.33\linewidth]{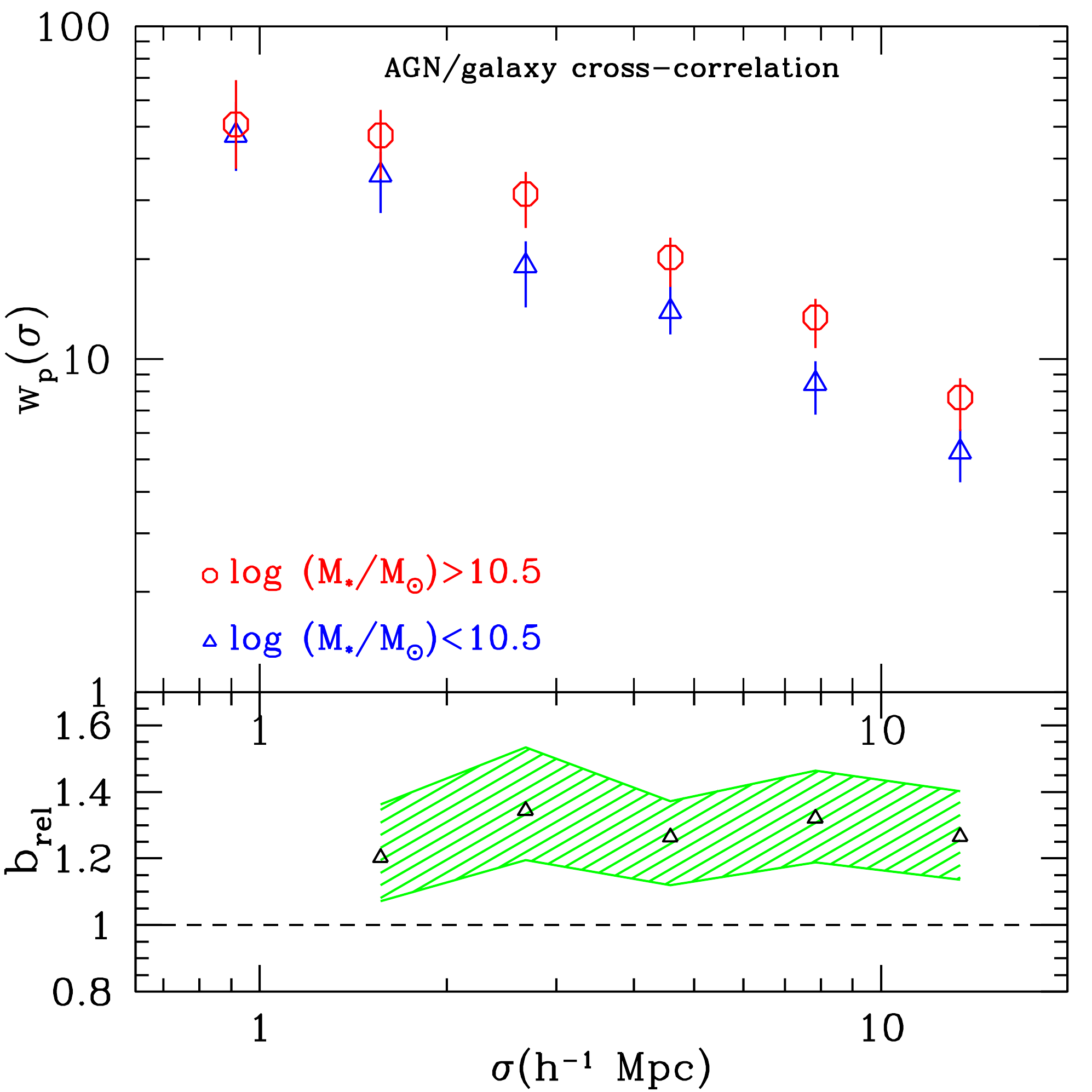}
\end{subfigure}%
\begin{subfigure}
  \centering
  \includegraphics[width=.33\linewidth]{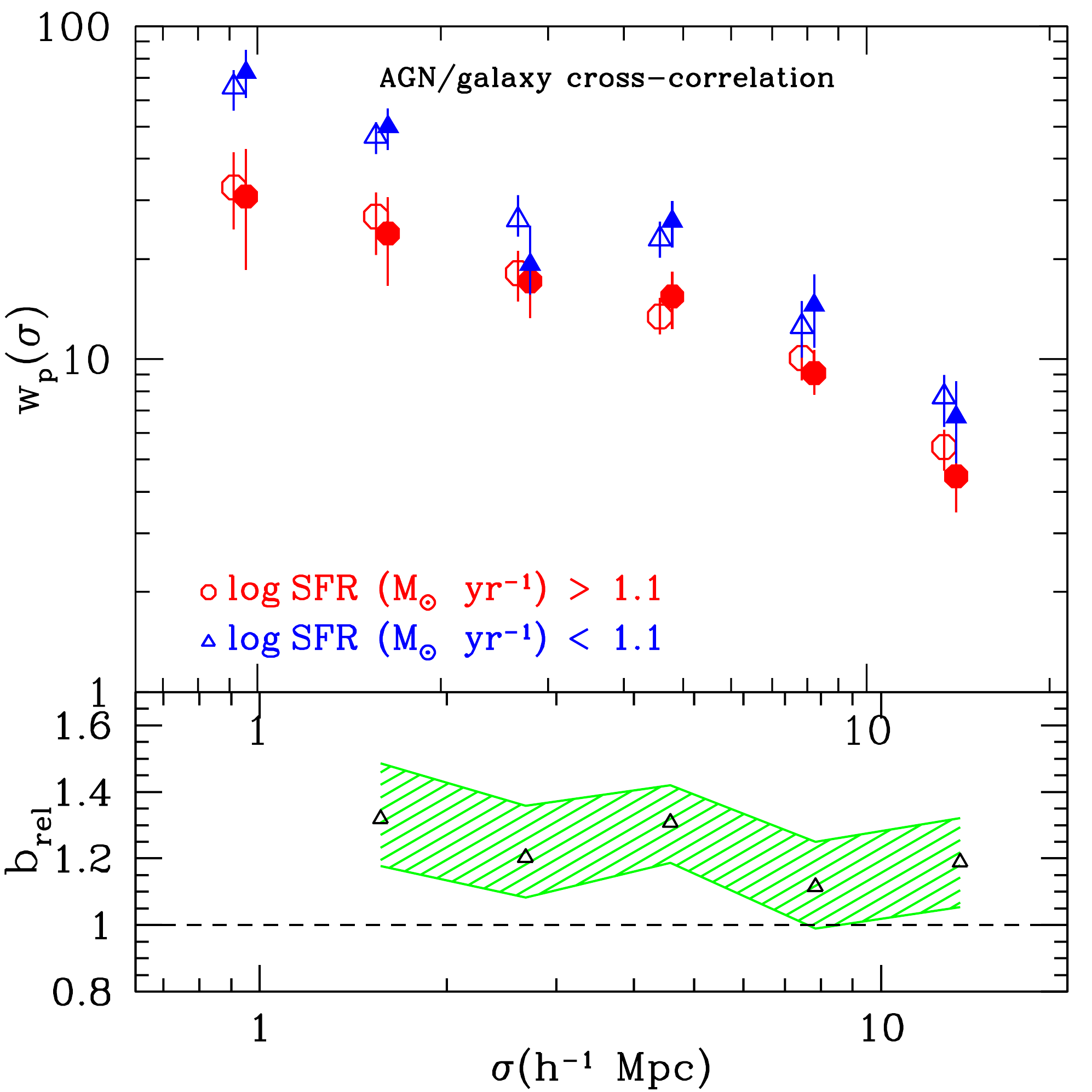}
\end{subfigure}%
\begin{subfigure}
  \centering
  \includegraphics[width=.33\linewidth]{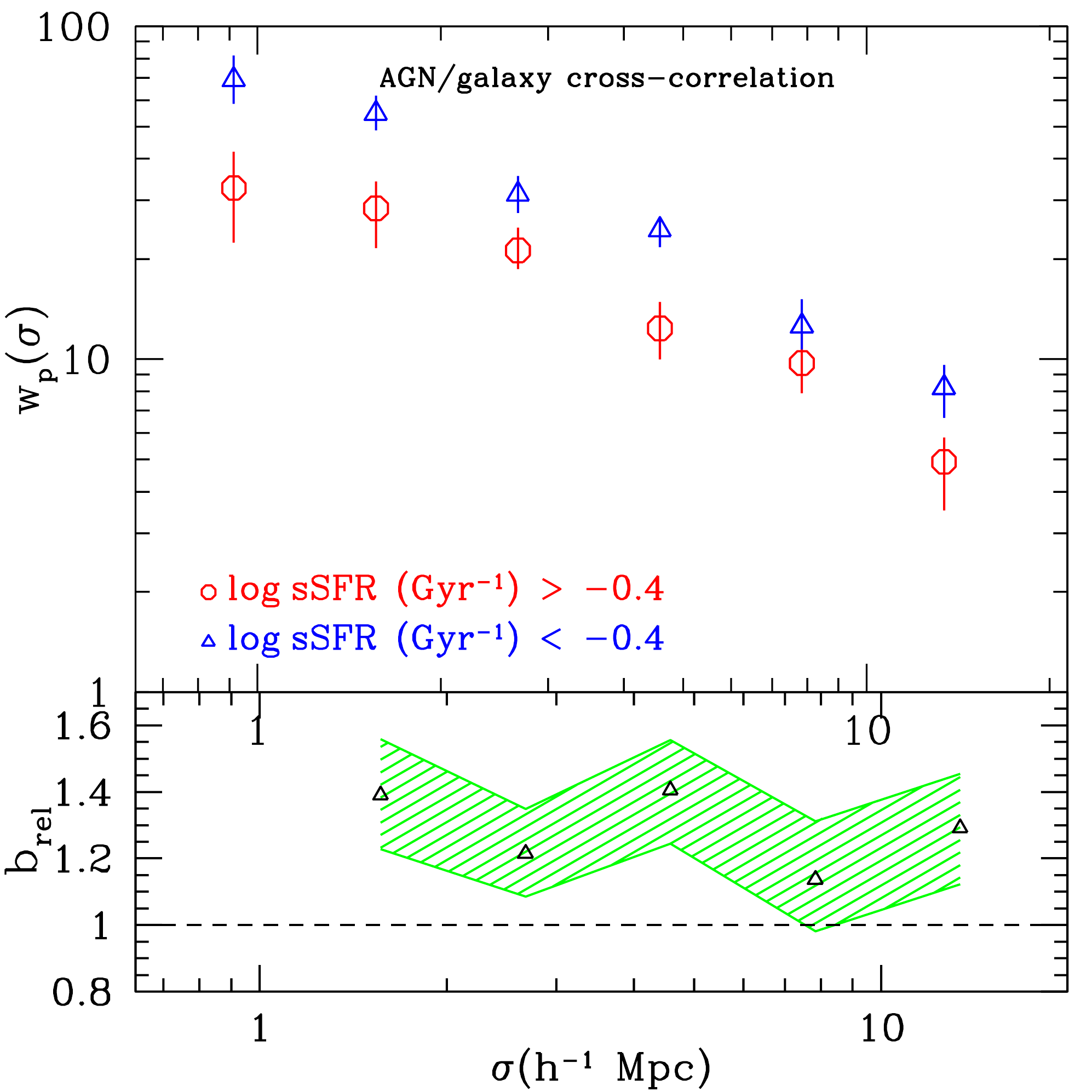}
\end{subfigure}%
\caption{{\bf{Left panel:}} The upper panel presents the AGN/galaxy cross-correlation functions for AGN with M$_{\star}>0.5$ (open circles) and M$_{\star}<0.5$ (open triangles). AGN hosted by more massive galaxies appear to have higher clustering amplitude compared to AGN hosted by less massive systems. The relative bias estimations as a function of separation, $\sigma$, are presented in the bottom panel. {\bf{Middle panel:}} The AGN/galaxy cross-correlation functions for AGN with $\log$ SFR$>1.1$ (open circles) and $\log$ SFR$<1.1$ (open triangles), are presented in the upper panel. Filled symbols present the measurements when we restrict the L$_X$ and redshift of the X-ray sample (see text for more details). The measurements reveal a negative X-ray clustering dependence on SFR, i.e., the higher the SFR the less clustered are the X-ray AGN. The bottom panel shows that relative bias estimations as a function of the separation. {\bf{Right panel:}} The AGN/galaxy cross-correlation functions for AGN with $\log$ sSFR\,$>-0.4$ (open circles) and $\log$ sSFR\,$<-0.4$ (open triangles) are presented in the top panel. A negative dependence of the X-ray clustering on sSFR is detected. The lower panel presents the relative bias calculations.}
\label{fig_clus_dependence}
\end{figure*}

We, also, created matched AGN and galaxy samples and compared their clustering. The correlation function measurements presented in Fig. \ref{fig_matched} show that the two populations have similar clustering. However, our measurements and those from previous clustering studies \citep[e.g.][]{Mendez2016}, limited by the small size of X-ray samples, study only the 2-halo term of the correlation function, i.e., $\sigma >1$h$^{-1}$\,Mpc and neglect the non-linear regime. Albeit, the 1-halo term can put constraints on e.g. the satellite halo occupation distribution (HOD) of AGN \citep[e.g.][]{Allevato2012} that will be compared to the satellite HOD of galaxies. Therefore, larger X-ray samples are required that will allow to incorporate the 1-halo term in the clustering analysis and examine whether the presence of an active SMBH affects the galaxy properties \citep[e.g.][]{Leauthaud2015}, especially at smaller scales.

Performing a similar analysis with that for the X-ray sources, we find that the galaxy clustering has similar dependence with the X-ray clustering on the properties of the (host) galaxy. In Fig. \ref{rel_bias_all}, we compare the relative bias of the AGN (green shaded region) and galaxies (blue shaded region) as a function of separation, when studying the M$_\star$ (top panel), SFR (middle panel) and sSFR (bottom panel) dependence of clustering. In all cases, active and normal galaxies present a similar dependence on the galaxy properties. However, the galaxy measurements have significantly smaller uncertainties compared to those for AGN. These results indicate that the existence of an active SMBH does no seem to affect the clustering properties of the galaxy it lives in. Previous clustering studies that used X-ray and optically-selected AGN found a weak clustering dependence on black hole mass and no significant dependence on Eddington ratio \citep{Krumpe2015}. Therefore, the AGN clustering seems to be (nearly) independent of the AGN properties, but it depends on the properties of the host galaxy.

Our findings are in agreement with most of the previous AGN and galaxy clustering studies. \cite{Mendez2016} used X-ray, radio and mid-IR AGN at $0.2<z<1.2$ from the PRIMUS and DEEP2 redshift surveys. They compared the clustering of each AGN sample with matched galaxy samples and found no statistical differences in their clustering properties. \cite{Meneux2009} used the zCOSMOS  galaxy sample with $9\leq \log (M_{\star} / M_{\odot}) \leq 11$ and found a mild dependence of the galaxy clustering on stellar mass up to $z=1$. \cite{Mostek2013} used DEEP2 galaxies to study the clustering dependence on stellar mass, SFR and sSFR, for star-forming and quiescent galaxies. They found a strong positive correlation on stellar mass for the blue galaxies but no dependence for the red ones. The stellar mass ranges they used for the two population, though, were different, i.e. $9.5\leq \log (M_{\star} / M_{\odot}) \leq 11.0$ for the star-forming and $10.5\leq \log (M_{\star} / M_{\odot}) \leq 11.5$ for the quiescent galaxies. They also found a positive correlation of the blue galaxy clustering with SFR and a negative correlation with sSFR. No correlation is found for the red population. However, when they constructed stellar mass-limited samples they found that the clustering amplitude increases with decreasing SFR, in agreement with our results. \cite{Coil2017} used a galaxy sample of 100,000 galaxies from the PRIMUS and DEEP2 surveys and found that the galaxy clustering is a stronger function of sSFR compared to stellar mass, in agreement with our findings for a strong (AGN and) galaxy clustering dependence on sSFR. Based on these results, \cite{Coil2017} suggest that knowing the halo mass of a galaxy is not sufficient to predict its stellar mass. The sSFR of the galaxy is also required. They attribute part of the scatter found in the stellar mass to halo mass relation \citep[e.g.][]{Behroozi2013} to the sSFR dependence of the halo mass. This finding contradicts halo models of galaxy evolution and favours the age-matching class of models. Our measurements support this idea and show that this picture holds regardless of the existence of an active SMBH in the centre of the galaxy.

\begin{figure}
\begin{center}
\includegraphics[height=1.\columnwidth]{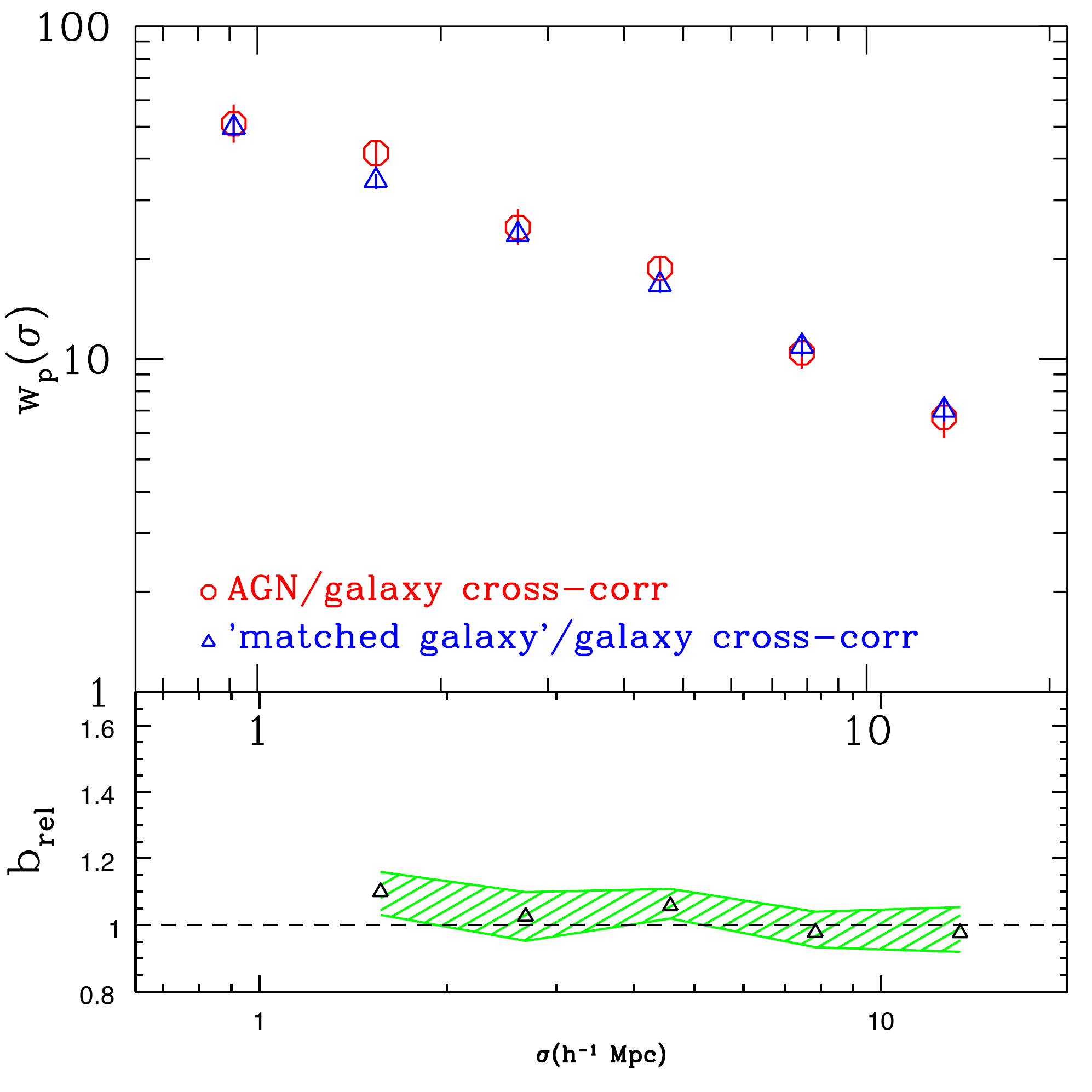}
\end{center} 
\caption{The AGN/galaxy (circles) and the matched galaxy/galaxy (triangles) cross-correlation functions. X-ray AGN and galaxies with matched $M_{\star}$, SFR and redshift distributions have similar clustering.}
\label{fig_matched}
\end{figure}

\begin{figure}
\begin{center}
\includegraphics[height=1.\columnwidth]{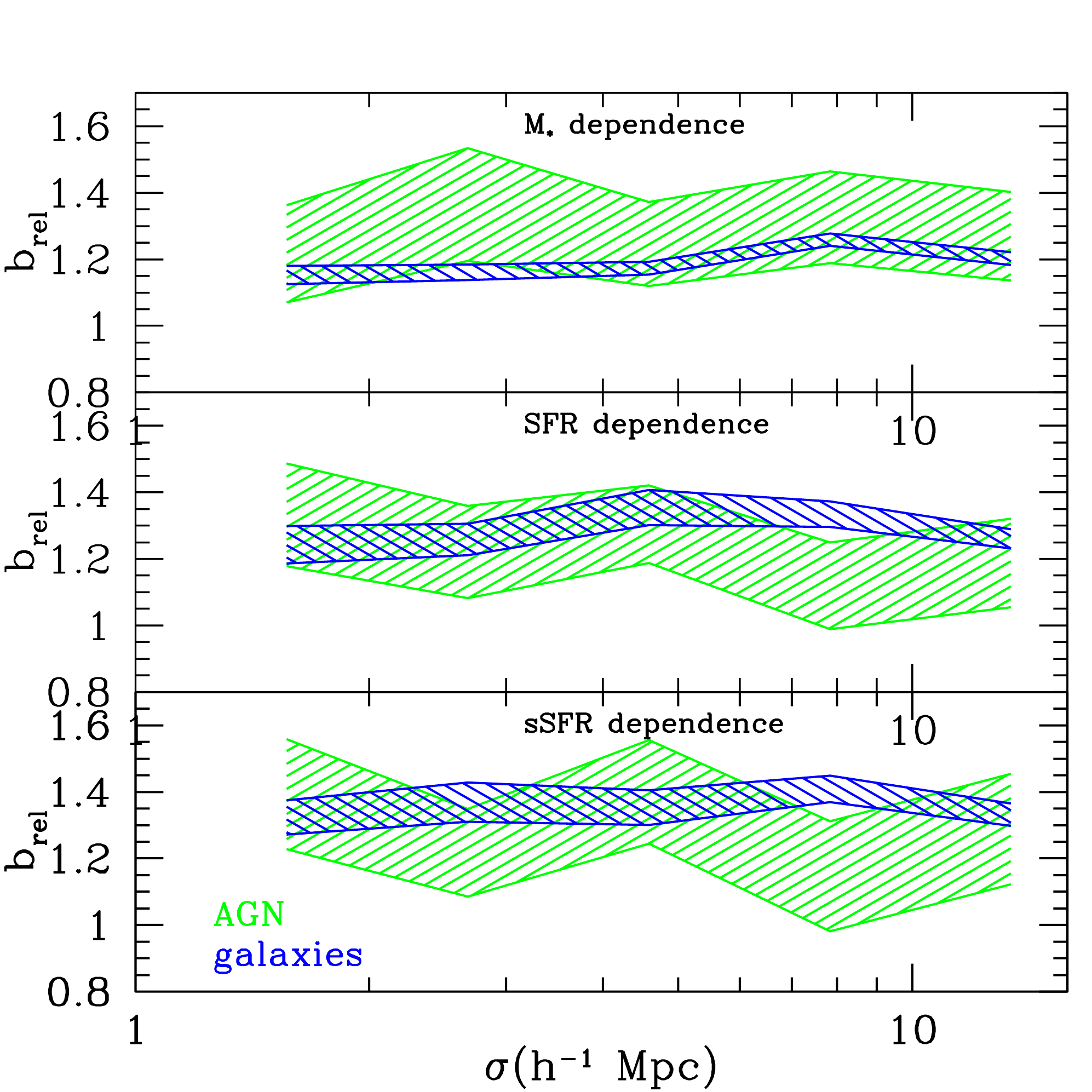}
\end{center} 
\caption{Relative bias, b$_{rel}$, estimations as a function of separation. The blue and green shaded regions present the measurements for the galaxies and the X-ray AGN, respectively. The top panel shows the dependence of the relative bias on stellar mass, the middle panel on SFR and the bottom panel on sSFR.}
\label{rel_bias_all}
\end{figure}

\section{Summary and Conclusions}
\label{sec_summary}

We have used the PDR-2 of the VIPERS survey to cross-correlate X-ray AGN from the XMM-XXL field with the VIPERS galaxies. The DMH mass estimations of the current study agree with the measurements of \cite{Mountrichas2016} that used the PDR-1 VIPERS dataset. 

The aim of this work is to study the dependence of clustering of X-ray AGN and normal galaxies on host galaxy properties, at $z\approx 0.8$. For that purpose, we matched the AGN and galaxy samples to have the same  $M_{\star}$, SFR and redshift distributions. Our results show that the two populations live in similar environments and their clustering has similar dependence on galaxy properties. Specifically, we find a positive dependence of the clustering on stellar mass and a negative dependence on SFR and  sSFR. Furthermore, the AGN clustering is independent of the location of the host galaxy above or below the star-forming main sequence.

X-ray AGN are hosted in a variety of environments. They live in galaxies that their properties, i.e., M$_{\star}$, SFR, redshift and X-ray luminosity, cover a wide baseline. To disentangle all these parameters and derive statistically robust measurements about the dependence of the AGN environment on each one of these host galaxy properties, requires X-ray samples orders of magnitude larger than those currently available. The 3XMM catalogue, under the condition of redshift availability for all its sources, could move us a step forward towards this direction. Future surveys, like eROSITA and ATHENA will allow us to exploit the large-scale environment of X-ray AGN in all their parameter space and uncover its dependence on all the aforementioned host galaxy properties.

\section{Acknowledgments}
The authors are grateful to the anonymous referee for helpful comments.ΠGM acknowledges financial support from the AHEAD project, which is funded by the European Union as Research and Innovation Action under Grant No: 654215.

\bibliography{mybib}{}

\begin{thebibliography}{63}
\expandafter\ifx\csname natexlab\endcsname\relax\def\natexlab#1{#1}\fi

\bibitem[{{Allevato} {et~al.}(2012)}]{Allevato2012}
{Allevato} V., {et~al.}, 2012, ApJ, 758, 47

\bibitem[{Behroozi {et~al.}(2013)Behroozi, Wechsler, \& Conroy}]{Behroozi2013}
Behroozi P.~S., Wechsler R.~H., Conroy C., 2013, ApJ, 770, 36

\bibitem[{{Bonoli} {et~al.}(2009){Bonoli}, {Marulli}, {Springel}, {White},
  {Branchini}, \& {Moscardini}}]{Bonoli2009}
{Bonoli} S., {Marulli} F., {Springel} V., {White} S.~D.~M., {Branchini} E.,
  {Moscardini} L., 2009, MNRAS, 396, 423

\bibitem[{Boyle \& Terlevich(1998)}]{Boyle1998}
Boyle B.~J., Terlevich R.~J., 1998, MNRAS, 293, 49

\bibitem[{Bruzual \& Charlot(2003)}]{Bruzual_Charlot2003}
Bruzual G., Charlot S., 2003, MNRAS, 344, 1000

\bibitem[{{Calzetti} {et~al.}(2000){Calzetti}, {Armus}, {Bohlin}, {Kinney},
  {Koornneef}, \& {Storchi-Bergmann}}]{Calzetti2000}
{Calzetti} D., {Armus} L., {Bohlin} R.~C., {Kinney} A.~L., {Koornneef} J.,
  {Storchi-Bergmann} T., 2000, ApJ, 533, 682

\bibitem[{{Cappelluti} {et~al.}(2012){Cappelluti}, {Allevato}, \&
  {Finoguenov}}]{Cappelluti2012}
{Cappelluti} N., {Allevato} V., {Finoguenov} A., 2012, ArXiv e-prints,
  1201.3920

\bibitem[{Casey(2012)}]{Casey2012}
Casey C.~M., 2012, MNRAS, 425, 3094

\bibitem[{Charlot \& Fall(2000)}]{Charlot_Fall_2000}
Charlot S., Fall S.~M., 2000, ApJ, 539, 718

\bibitem[{{Ciesla} {et~al.}(2015)}]{Ciesla2015}
{Ciesla} L., {et~al.}, 2015, A\&A, 576, 19

\bibitem[{{Coil} {et~al.}(2017){Coil}, {Mendez}, {Eisenstein}, \&
  {Moustakas}}]{Coil2017}
{Coil} A.~L., {Mendez} A.~J., {Eisenstein} D.~J., {Moustakas} J., 2017, ApJ,
  838, 87

\bibitem[{{Coil} {et~al.}(2009)}]{Coil2009}
{Coil} A.~L., {et~al.}, 2009, ApJ, 701, 1484

\bibitem[{{Croom} {et~al.}(2004){Croom}, {Smith}, {Boyle}, {Shanks}, {Miller},
  {Outram}, \& {Loaring}}]{Croom2004}
{Croom} S.~M., {Smith} R.~J., {Boyle} B.~J., {Shanks} T., {Miller} L., {Outram}
  P.~J., {Loaring} N.~S., 2004, MNRAS, 349, 1397

\bibitem[{{Dale} {et~al.}(2014){Dale}, {Helou}, {Magdis}, {Armus},
  {D{\'{\i}}az-Santos}, \& {Shi}}]{Dale2014}
{Dale} D.~A., {Helou} G., {Magdis} G.~E., {Armus} L., {D{\'{\i}}az-Santos} T.,
  {Shi} Y., 2014, ApJ, 784, 83

\bibitem[{Dawson {et~al.}(2013)}]{Dawson2013}
Dawson K.~S., {et~al.}, 2013, AJ, 145, 10

\bibitem[{de~la Torre {et~al.}(2017)}]{delaTorre2017}
de~la Torre S., {et~al.}, 2017, A\&A, 608, 44

\bibitem[{Draine \& Li(2007)}]{Draine_Li_2007}
Draine B.~T., Li A., 2007, ApJ, 657, 810

\bibitem[{{Eisenstein} {et~al.}(2011)}]{Eisenstein2011}
{Eisenstein} D.~J., {et~al.}, 2011, AJ, 142, 72

\bibitem[{{Fanidakis} {et~al.}(2013)}]{Fanidakis2013a}
{Fanidakis} N., {et~al.}, 2013, MNRAS, 435, 679

\bibitem[{{Ferrarese} \& {Merritt}(2000)}]{Ferrarese2000}
{Ferrarese} L., {Merritt} D., 2000, ApJ, 539, 9

\bibitem[{{Fotopoulou} {et~al.}(2016)}]{Fotopoulou2016}
{Fotopoulou} S., {et~al.}, 2016, A\&A, 587, 16

\bibitem[{{Fritz} {et~al.}(2006){Fritz}, {Franceschini}, \&
  {Hatziminaoglou}}]{Fritz2006}
{Fritz} J., {Franceschini} A., {Hatziminaoglou} E., 2006, MNRAS, 166, 767

\bibitem[{Garilli {et~al.}(2014)}]{Garilli2014}
Garilli B., {et~al.}, 2014, A\&A, 562, 23

\bibitem[{{Georgakakis} {et~al.}(2014)}]{Georgakakis2014}
{Georgakakis} A., {et~al.}, 2014, MNRAS, 443, 3327

\bibitem[{Georgakakis {et~al.}(2017)}]{Georgakakis2017}
Georgakakis A., {et~al.}, 2017, MNRAS, 469, 3232

\bibitem[{Guzzo {et~al.}(2014)}]{Guzzo2014}
Guzzo L., {et~al.}, 2014, A\&A, 566, 108

\bibitem[{{Hopkins} {et~al.}(2008){Hopkins}, {Hernquist}, {Cox}, \&
  {Keres}}]{Hopkins2008a}
{Hopkins} P.~F., {Hernquist} L., {Cox} T.~J., {Keres} D., 2008, ApJS, 175, 356

\bibitem[{{Hopkins} {et~al.}(2006){Hopkins}, {Hernquist}, {Cox}, {Robertson},
  {Di Matteo}, \& {Springel}}]{Hopkins2006a}
{Hopkins} P.~F., {Hernquist} L., {Cox} T.~J., {Robertson} B., {Di Matteo} T.,
  {Springel} V., 2006, ApJ, 639, 700

\bibitem[{Kormendy(2013)}]{Kormendy2013}
Kormendy J. \&~Ho L.~C., 2013, ARAA, 51, 511

\bibitem[{Koutoulidis {et~al.}(2013)Koutoulidis, Plionis, Georgantopoulos, \&
  Fanidakis}]{Koutoulidis2013}
Koutoulidis L., Plionis M., Georgantopoulos I., Fanidakis N., 2013, MNRAS, 428,
  1382

\bibitem[{{Krumpe} {et~al.}(2010){Krumpe}, {Miyaji}, \& {Coil}}]{krumpe2010}
{Krumpe} M., {Miyaji} T., {Coil} A.~L., 2010, ApJ, 713, 558

\bibitem[{{Krumpe} {et~al.}(2012){Krumpe}, {Miyaji}, {Coil}, \&
  {Aceves}}]{Krumpe2012}
{Krumpe} M., {Miyaji} T., {Coil} A.~L., {Aceves} H., 2012, ApJ, 746, 1

\bibitem[{{Krumpe} {et~al.}(2018){Krumpe}, {Miyaji}, {Coil}, \&
  {Hector}}]{Krumpe2018}
{Krumpe} M., {Miyaji} T., {Coil} A.~L., {Hector} A., 2018, MNRAS, 474, 1773

\bibitem[{{Krumpe} {et~al.}(2015){Krumpe}, {Miyaji}, {Husemann}, {Fanidakis},
  {Coil}, \& {Aceves}}]{Krumpe2015}
{Krumpe} M., {Miyaji} T., {Husemann} B., {Fanidakis} N., {Coil} A.~L., {Aceves}
  H., 2015, ApJ, 815, 21

\bibitem[{{Le F{\`e}vre} {et~al.}(2003){Le F{\`e}vre}, {Saisse}, {Mancini},
  {Brau-Nogue}, {Caputi}, {Castinel}, {D'Odorico}, {Garilli}, {Kissler-Patig},
  {Lucuix}, {Mancini}, {Pauget}, {Sciarretta}, {Scodeggio}, {Tresse}, \&
  {Vettolani}}]{LeFevre2003}
{Le F{\`e}vre} O., {Saisse} M., {Mancini} D., {Brau-Nogue} S., {Caputi} O.,
  {Castinel} L., {D'Odorico} S., {Garilli} B., {Kissler-Patig} M., {Lucuix} C.,
  {Mancini} G., {Pauget} A., {Sciarretta} G., {Scodeggio} M., {Tresse} L.,
  {Vettolani} G., 2003, in Society of Photo-Optical Instrumentation Engineers
  (SPIE) Conference Series, Vol. 4841, Instrument Design and Performance for
  Optical/Infrared Ground-based Telescopes, {Iye} M., {Moorwood} A.~F.~M.,
  eds., pp. 1670--1681

\bibitem[{{Leauthaud} {et~al.}(2015)}]{Leauthaud2015}
{Leauthaud} A., {et~al.}, 2015, MNRAS, 446, 1874

\bibitem[{Liu {et~al.}(2016)Liu, Merloni, Georgakakis, Menzel, Buchner, Nandra,
  Salvato, Shen, Brusa, \& Streblyanska}]{Liu2016}
Liu Z., Merloni A., Georgakakis A., Menzel M.-L., Buchner J., Nandra K.,
  Salvato M., Shen Y., Brusa M., Streblyanska A., 2016, MNRAS, 459, 1602

\bibitem[{{Lusso} {et~al.}(2012)}]{Lusso2012}
{Lusso} E., {et~al.}, 2012, MNRAS, 425, 623

\bibitem[{Magorrian {et~al.}(1998)}]{Magorrian2008}
Magorrian J., {et~al.}, 1998, AJ, 115, 2285

\bibitem[{{Marulli} {et~al.}(2009){Marulli}, {Bonoli}, {Branchini}, {Gilli},
  {Moscardini}, \& {Springel}}]{Marulli2009}
{Marulli} F., {Bonoli} S., {Branchini} E., {Gilli} R., {Moscardini} L.,
  {Springel} V., 2009, MNRAS, 396, 1404

\bibitem[{Masoura {et~al.}(2018)Masoura, Mountrichas, Georgantopoulos, Ruiz,
  Magdis, \& Plionis}]{Masoura2018}
Masoura V.~A., Mountrichas G., Georgantopoulos I., Ruiz A., Magdis G., Plionis
  M., 2018, eprint arXiv:1807.01723

\bibitem[{{Mendez} {et~al.}(2016)}]{Mendez2016}
{Mendez} A.~J., {et~al.}, 2016, ApJ, 821, 55

\bibitem[{{Meneux} {et~al.}(2009)}]{Meneux2009}
{Meneux} B., {et~al.}, 2009, A\&A, 505, 463

\bibitem[{Menzel {et~al.}(2016)Menzel, Merloni, Georgakakis, Salvato, Aubourg,
  Brandt, Brusa, Buchner, Dwelly, Nandra, P{\^a}ris, Petitjean, \&
  Schwope}]{Menzel2016}
Menzel M.-L., Merloni A., Georgakakis A., Salvato M., Aubourg E., Brandt W.~N.,
  Brusa M., Buchner J., Dwelly T., Nandra K., P{\^a}ris I., Petitjean P.,
  Schwope A., 2016, MNRAS, 457, 110

\bibitem[{{Miyaji} {et~al.}(2011){Miyaji}, {Krumpe}, {Coil}, \&
  {Aceves}}]{Miyaji2011}
{Miyaji} T., {Krumpe} M., {Coil} A.~L., {Aceves} H., 2011, ApJ, 726, 83

\bibitem[{{Mostek} {et~al.}(2013){Mostek}, {Coil}, {Cooper}, {Davis}, {Newman},
  \& {Weiner}}]{Mostek2013}
{Mostek} N., {Coil} A.~L., {Cooper} M., {Davis} M., {Newman} J.~A., {Weiner}
  B.~J., 2013, ApJ, 767, 89

\bibitem[{{Mountrichas} \& {Georgakakis}(2012)}]{Mountrichas2012}
{Mountrichas} G., {Georgakakis} A., 2012, MNRAS, 420, 514

\bibitem[{{Mountrichas} {et~al.}(2009){Mountrichas}, {Sawangwit}, {Shanks},
  {Croom}, P., {Myers}, \& {Pimbblet}}]{Mountrichas2009}
{Mountrichas} G., {Sawangwit} U., {Shanks} T., {Croom} S.~M., P. S.~D., {Myers}
  A.~D., {Pimbblet} K., 2009, MNRAS, 394, 2050

\bibitem[{{Mountrichas} {et~al.}(2016)}]{Mountrichas2016}
{Mountrichas} G., {et~al.}, 2016, MNRAS, 457, 4195

\bibitem[{Noeske {et~al.}(2007)}]{Noeske2007}
Noeske K.~G., {et~al.}, 2007, ApJL, 660, 43

\bibitem[{{Noll} {et~al.}(2009)}]{Noll2009}
{Noll} S., {et~al.}, 2009, A\&A, 507, 1793

\bibitem[{Pezzotta {et~al.}(2017)}]{Pezzotta2017}
Pezzotta A., {et~al.}, 2017, A\&A, 7, 18

\bibitem[{{Pierre} {et~al.}(2016)}]{Pierre2016}
{Pierre} M., {et~al.}, 2016, A\&A, 592, 1

\bibitem[{{Powell} {et~al.}(2018)}]{Powell2018}
{Powell} M., {et~al.}, 2018, arXiv:1803.07589

\bibitem[{{Pozzetti} {et~al.}(2010)}]{Pozzetti2010}
{Pozzetti} L., {et~al.}, 2010, A\&A, 523, 23

\bibitem[{Risaliti {et~al.}(2002)Risaliti, {Elvis}, \& {Gilli}}]{Risaliti2002}
Risaliti G., {Elvis} M., {Gilli} R., 2002, ApJL, 566, 67

\bibitem[{Schreiber {et~al.}(2015)}]{Schreiber2015}
Schreiber C., {et~al.}, 2015, A\&A, 575, 29

\bibitem[{{Scodeggio}(2016)}]{Scodeggio2016}
{Scodeggio} M., 2016, eprint arXiv:1611.07048

\bibitem[{{Sheth} {et~al.}(2001){Sheth}, {Mo}, \& {Tormen}}]{Sheth2001}
{Sheth} R.~K., {Mo} H.~J., {Tormen} G., 2001, MNRAS, 323, 1

\bibitem[{{Smee} {et~al.}(2013)}]{Smee2013}
{Smee} S., {et~al.}, 2013, AJ, 146, 32

\bibitem[{{Starikova} {et~al.}(2011)}]{Starikova2011}
{Starikova} S., {et~al.}, 2011, ApJ, 741, 15

\bibitem[{{Sutherland} \& {Saunders}(1992)}]{Sutherland_and_Saunders1992}
{Sutherland} W., {Saunders} W., 1992, MNRAS, 259, 413

\bibitem[{{Van den Bosch}(2002)}]{Bosch2002}
{Van den Bosch} F.~C., 2002, MNRAS, 331, 98

\end{thebibliography}
\bibliographystyle{mn2e}

\clearpage

\appendix

\section{Using NIR photometry for the SED fitting}
\label{append_nir}

The AGN and galaxy SEDs we fit with CIGALE to estimate stellar mass, have been constructed using optical photometry (CFHTLS). In this Section we test whether adding NIR photometry in the SED fitting process affects the stellar mass calculations.

To add NIR photometry to the X-ray and galaxy samples, we match the datasets with NIR datasets (VHS-DR5 and VIKING) using a 5'' radius. Only sources with good photometry in at least one of the NIR bands and with low probability of being instrumental noise (PNOISE$<$0.05) were selected. After all selection criteria are applied (see Sections \ref{agn_samples} and \ref{galaxy_samples}) there are 239 AGN and 16,643 galaxies, with NIR photometry available. Figures \ref{agn_mstar_nir} and \ref{gal_mstar_nir} present the stellar mass distributions for AGN and galaxies, respectively, with (blue lines) and without (red lines) NIR photometry in the SED fitting analysis. The inclusion of NIR photometry in the SED fitting analysis, gives a small but systematic difference in the stellar mass estimations ($<0.25$\,dex) compared to M$_\star$ calculations without including near-IR photometry, for both the AGN and galaxy samples. This small difference is within the typical errors of mass estimations due to e.g. the different templates chosen in the SED analysis (see Appendix \ref{append_SED_libraries}).

Therefore, dropping the requirement for NIR photometry does not affect our clustering analysis and increases the number of available sources used in our measurements.

\begin{figure*}
\centering
\begin{subfigure}
  \centering
  \includegraphics[width=.45\linewidth]{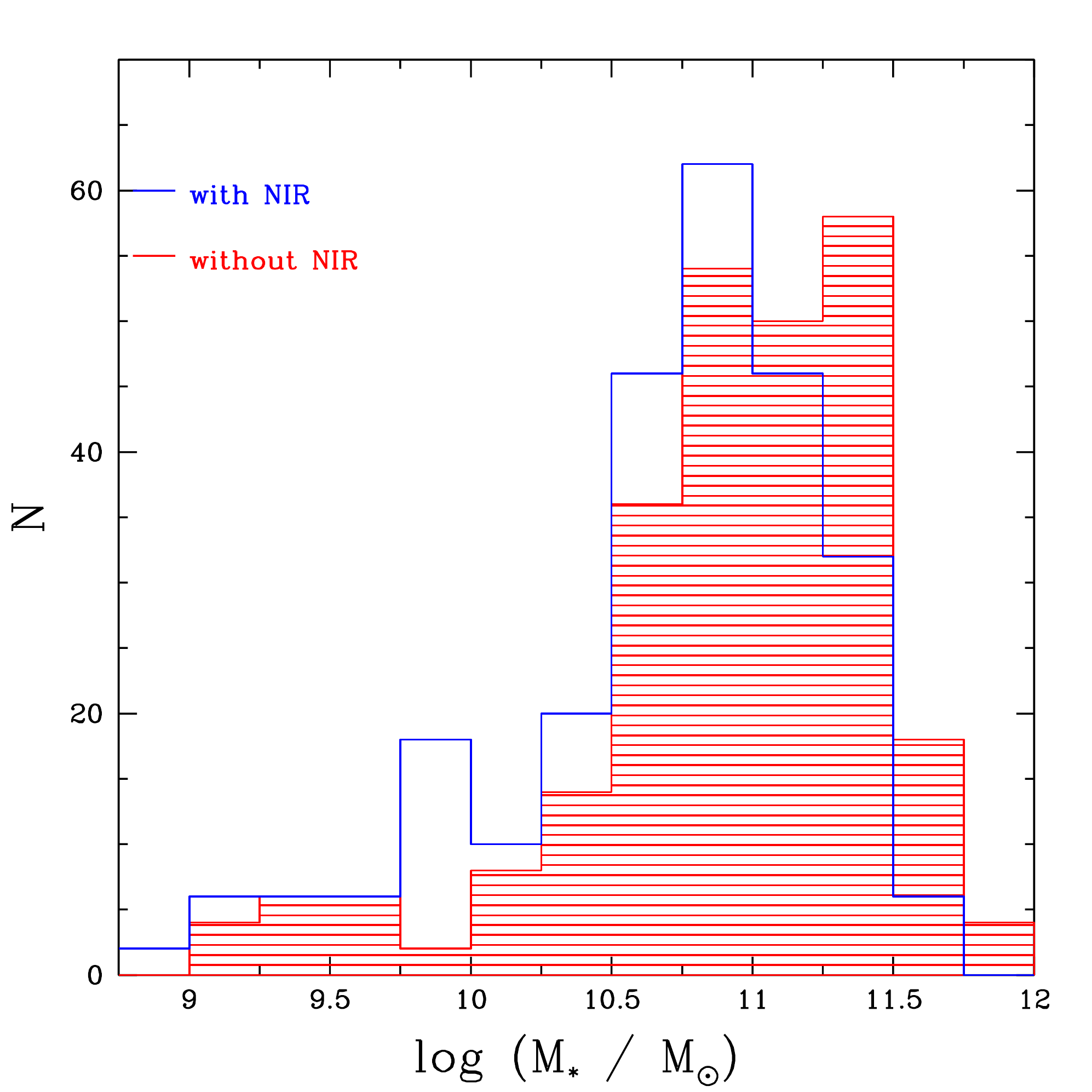}
\end{subfigure}
\begin{subfigure}
  \centering
  \includegraphics[width=.45\linewidth]{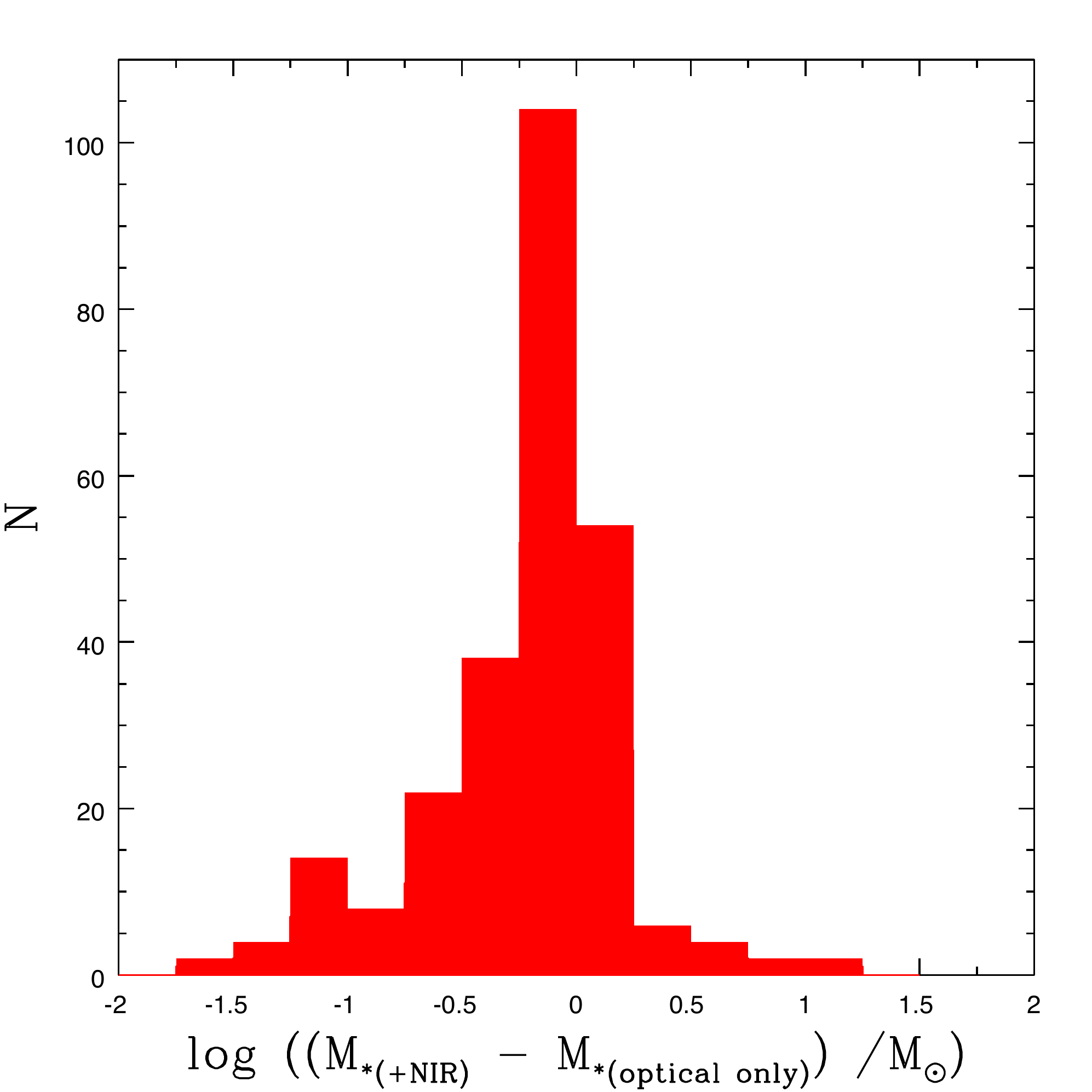}
\end{subfigure}
\caption{Left: The AGN stellar mass using CFHTLS optical photometry alone (red shaded area) and additionally NIR photometry (blue line). Right: The distribution of the M$_{\star, \rm{+NIR}}$$-$M$_{\star, \rm{optical\,only}}$. There is a $<0.25$\,dex difference in the M$_\star$ estimations when adding NIR photometry to the AGN SED fitting analysis.}
\label{agn_mstar_nir}
\end{figure*}

\begin{figure*}
\centering
\begin{subfigure}
  \centering
  \includegraphics[width=.45\linewidth]{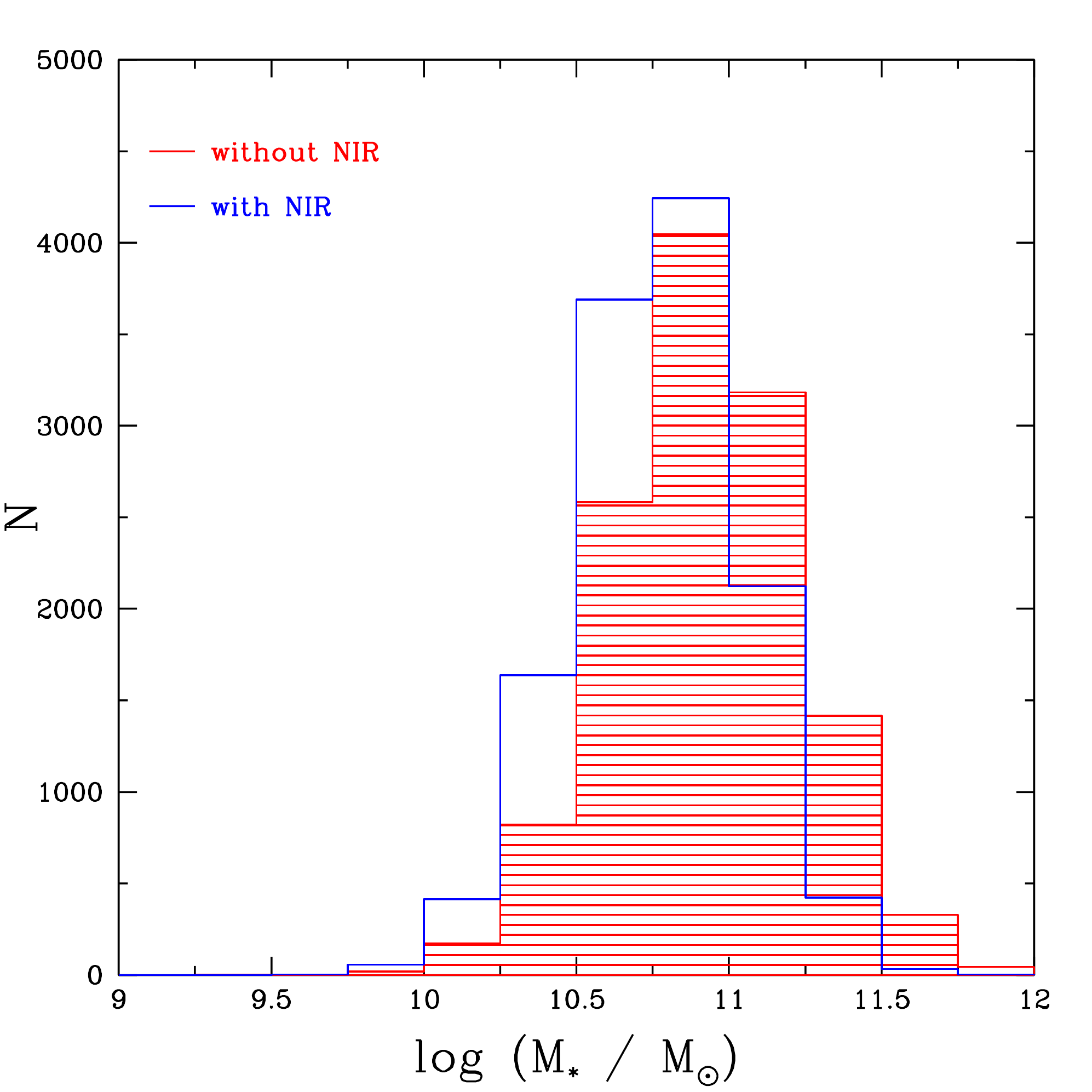}
\end{subfigure}
\begin{subfigure}
  \centering
  \includegraphics[width=.45\linewidth]{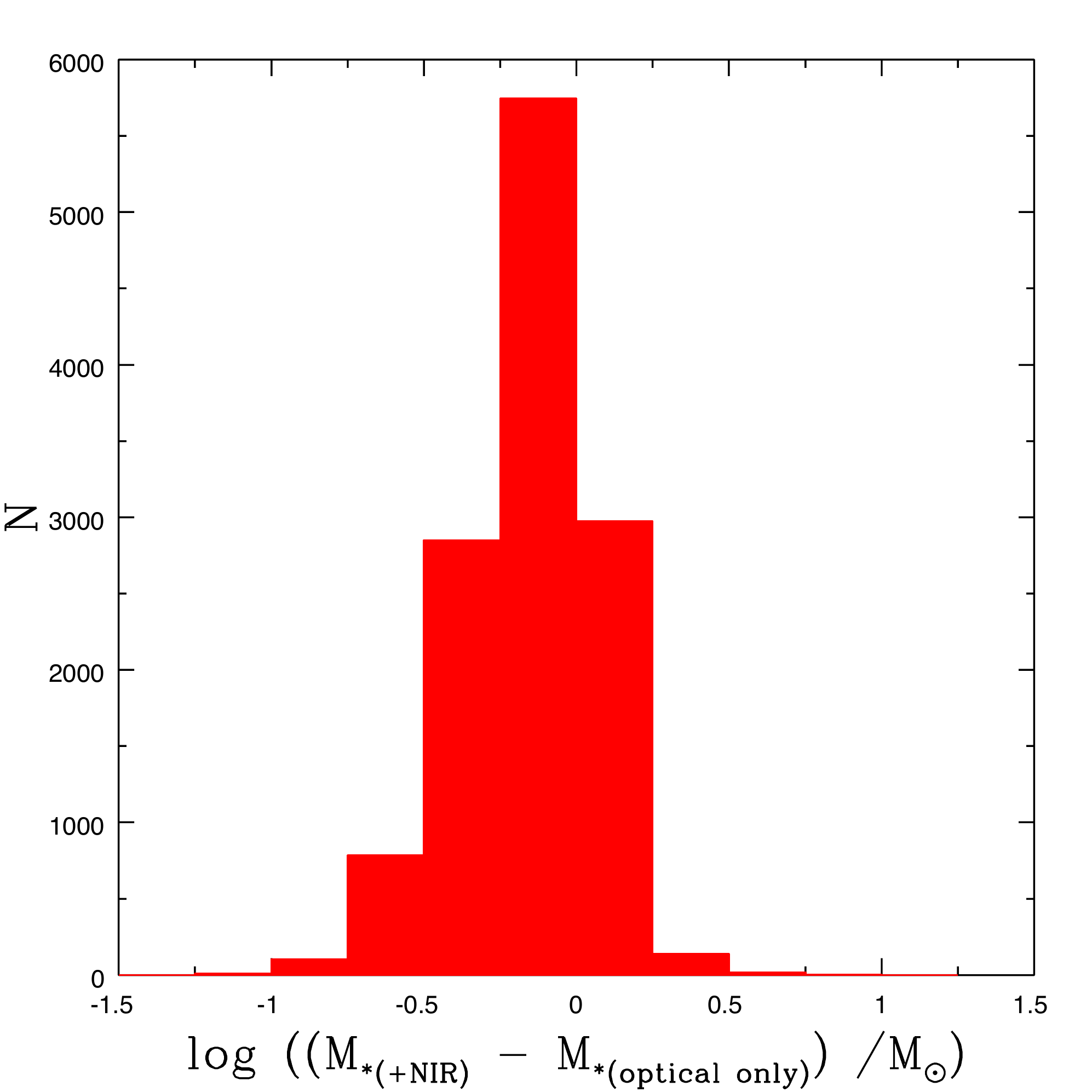}
\end{subfigure}
\caption{Left: The galaxy stellar mass using CFHTLS optical photometry alone (red shaded area) and additionally NIR photometry (blue line). Right: The distribution of the M$_{\star, \rm{+NIR}}$$-$M$_{\star, \rm{optical\,only}}$. There is a $<0.25$\,dex difference in the M$_\star$ estimations when adding NIR photometry to the galaxy SED fitting analysis.}
\label{gal_mstar_nir}
\end{figure*}

\section{AGN SED fitting: SDSS vs. CFHTLS photometry}
\label{append_SDSS_CFHTLS}

In our analysis we use CFHTLS, instead of SDSS, optical photometry for the SED fitting of the AGN sample. The justification is that the VIPERS galaxy sample is selected from the optical photometric catalogues of the CFHTLS. Therefore, this choice allow us to be consistent in the SED fitting analysis of the AGN and galaxy samples. The downside, however, is that it reduces the number of available X-ray sources, by $\sim10\%$. Specifically, 364 AGN in the X-ray sample have CFHTLS observations out of the 407 detected by SDSS (Section \ref{agn_samples}). If the choice of the optical photometry does not affect the SED measurements then we can use the largest X-ray sample in our analysis. Thus, we compare the SED results using the two different optical surveys.

Utilizing the models and the values for their free parameters presented in Table \ref{table_cigale}, we run CIGALE for the  305 AGN sample using SDSS photometry. Fig. \ref{agn_mstar_cfhtls_sdss} (left panel) compares the distribution of the stellar mass estimations with CFHTLS (solid line) and SDSS (dashed line) optical photometry. In the right panel we present the distribution of the difference in the two stellar mass distributions, i.e., $\log$\,((M$_{\star, \rm{CFHTLS}}$-M$_{\star, \rm{SDSS}})/M_\odot)$. The measurements show that there is a $\sim 0.25$\,dex difference in the M$_\star$ estimations using different optical photometry. Although this difference will not affect our AGN clustering dependence analysis on stellar mass, it will affect the estimated galaxy weights when we match the AGN and galaxy samples (see Section \ref{sec_galaxy_matched}). Therefore, we use CFHTLS optical photometry for the AGN SED fitting to be consistent with the galaxy stellar mass estimations.

\begin{figure*}
\centering
\begin{subfigure}
  \centering
  \includegraphics[width=.45\linewidth]{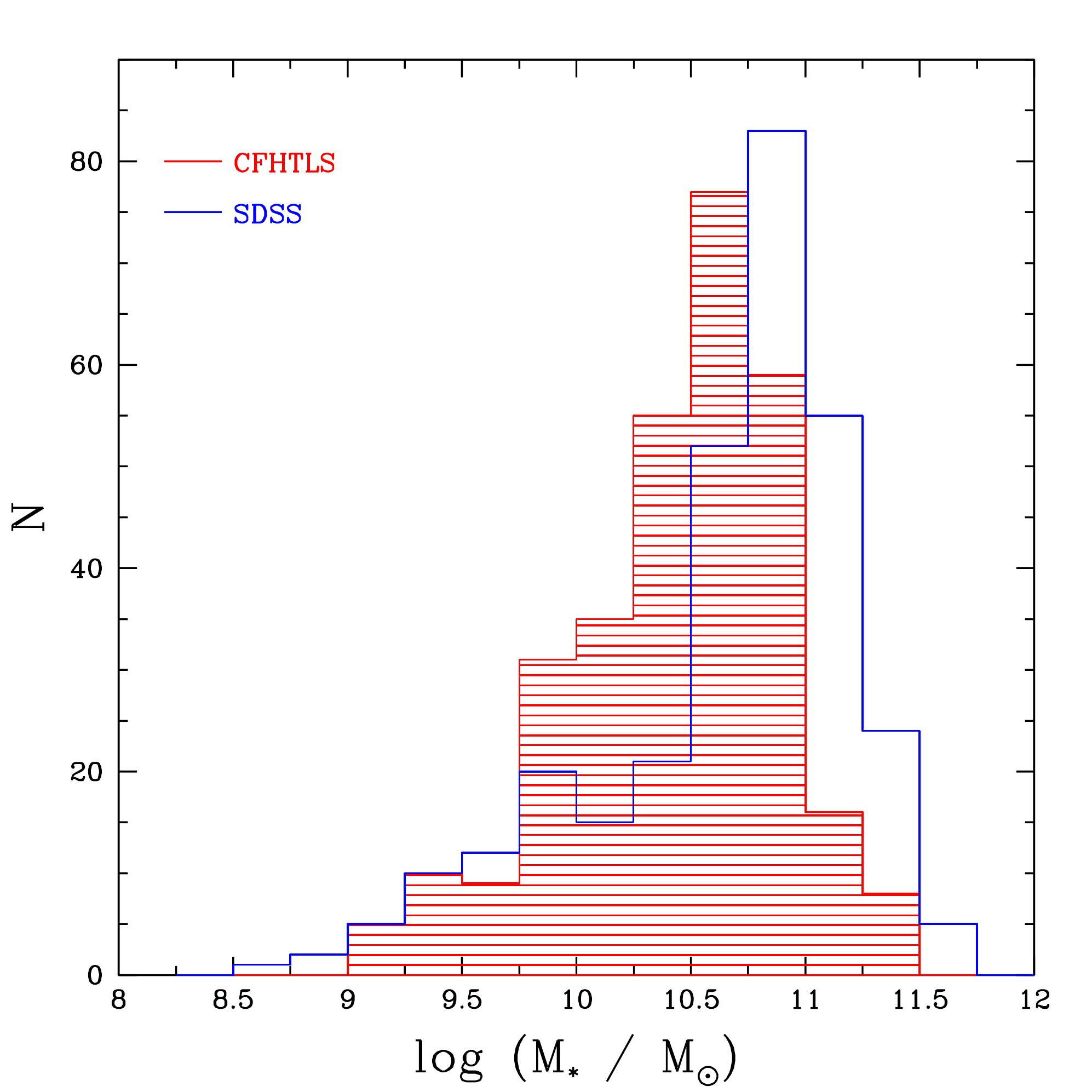}
\end{subfigure}
\begin{subfigure}
  \centering
  \includegraphics[width=.45\linewidth]{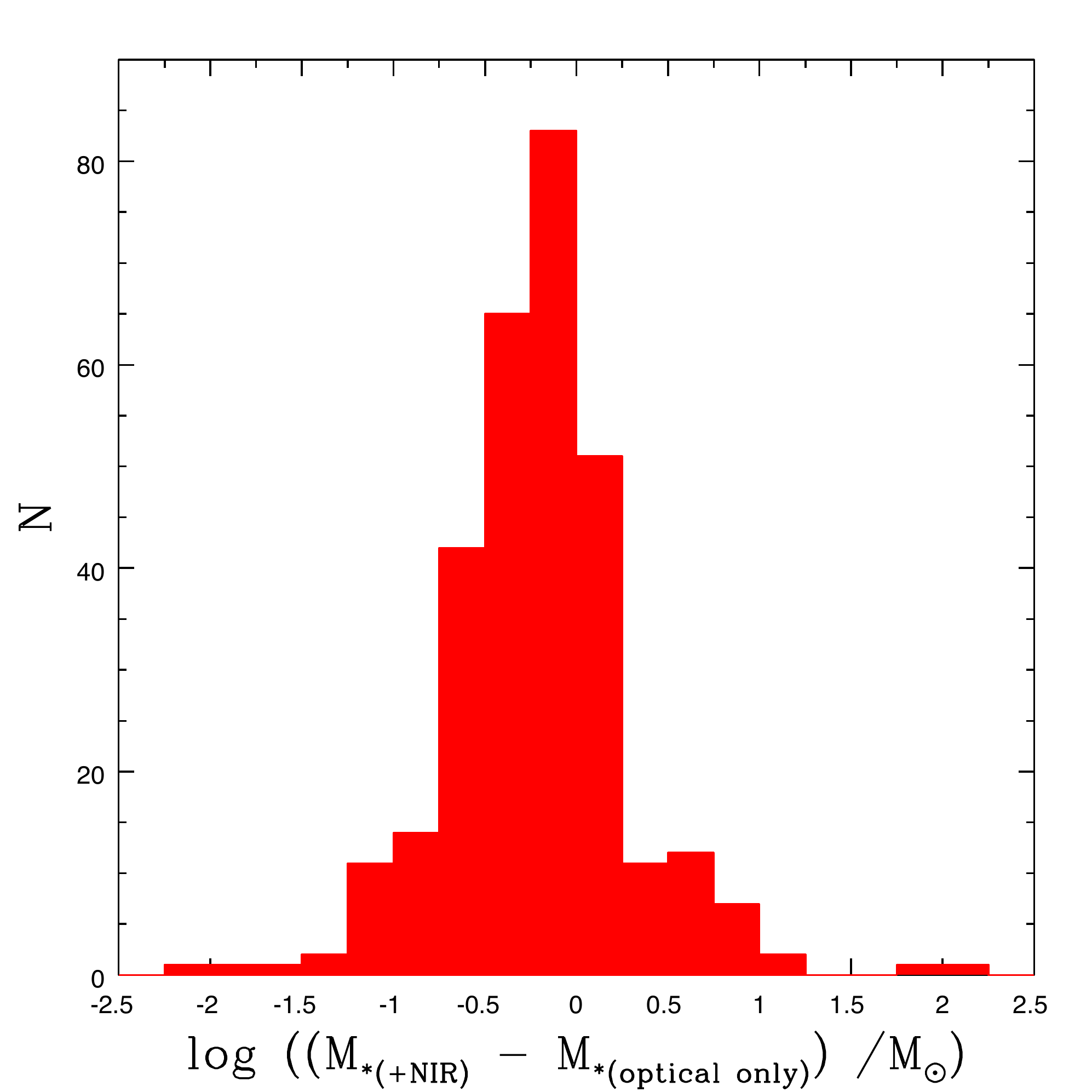}
\end{subfigure}
\caption{Left: The AGN stellar mass using CFHTLS optical photometry (red line) and SDSS photometry (blue line). Right: The distribution of the M$_{\star, \rm{CFHTLS}}$-M$_{\star, \rm{SDSS}}$. There is a $\sim 0.25$\,dex difference in the M$_\star$ estimations using different optical photometry.}
\label{agn_mstar_cfhtls_sdss}
\end{figure*}

\section{Sensitivity of the SED fitting measurements to the choice of different libraries}
\label{append_SED_libraries}

\begin{table*}
\caption{The mean values and the standard deviation, $\sigma_{sd}$,  for the M$_*$ , SFR and sSFR of the AGN sample, using different libraries for the SED fitting. Column 2 presents the measurements for the modules presented in Table 1 that are used in our clustering analysis.}
\centering
\setlength{\tabcolsep}{0.7mm}
\begin{tabular}{cccccccccc}
       \hline
& & SFH & \multicolumn{2}{c}{Dust attenuation} & \multicolumn{3}{c}{Dust Emission} & AGN templates\\
& modules in Table 1 & delayed & Charlot \& Fall (2000) & 2-powerlaws & Casey et al. (2012) & Draine \& Li (2007) & Draine \& Li (2014) & Dale et al. (2014) & mean$\pm \sigma_{sd}$\\
       \hline
SFR & 1.31 & 1.42 & 1.47& 1.53& 1.30& 1.35 & 1.32& 1.42& 1.39$\pm$0.08\\
M$_*$ & 10.65 & 10.65 & 10.80 & 10.89 & 10.64 & 10.71& 10.72 & 11.00 & 10.76$\pm$0.12\\
sSFR & -0.33 & -0.35 & -0.31 & -0.31 & -0.32 & -0.37& -0.39 & -0.63 & -0.37$\pm$0.08\\
\hline
\label{table:diff_modules}
\end{tabular}
\end{table*}

In the SED fitting measurements we use the libraries shown in Table \ref{table_cigale}. In this Section, we study how sensitive are the measurements to the choice of different libraries. Specifically, for the SFH, we use a delayed SFH model instead of the double-exponentially-decreasing (2-$\tau$) model. For the dust attenuation we replace the \cite{Calzetti2000} template with the \cite{Charlot_Fall_2000} single powerlaw attenuation module as well as a double powerlaw module. The double powerlaw implements an attenuation law combining the birth cloud attenuation and the interstellar medium attenuation, each one modelled by a power law. The dust emission is now modelled with \cite{Casey2012}, the \cite{Draine_Li_2007} and the updated  \cite{Draine_Li_2007} infrared models  instead of the \cite{Dale2014}. For the AGN we always use the \cite{Fritz2006} library. This library has been added and extensively tested to match CIGALE's philosophy. For more details on each module see \cite{Ciesla2015} and references therein. 

Fig. \ref{fig_diff_modules} presents the SFR, M${_*}$ and sSFR distributions derived using the various templates when fitting the AGN SEDs. Table \ref{table:diff_modules} presents the mean values for the aforementioned host galaxy properties. Using a delayed SFH template (dotted, blue line) produces lower SFR compared to the 2-$\tau$ model (solid, black line) However, the SFR distribution peaks at higher values and therefore the estimated mean SFR of the AGN sample is $\sim 0.1$\,dex higher when using the delayed SFH module. The SFR distribution peaks at $\sim 0.25$\,dex higher values when modelling the dust attenuation with either a single or a double powerlaw model, compared to the \cite{Calzetti2000} template. This translates into $\sim 0.20$\,dex higher mean SFR values. Using different dust emission templates does not seem to affect the SFR and M${_*}$ calculations. 

From the mean  SFR, M${_*}$ and sSFR values presented in Table \ref{table:diff_modules} we conclude that using different templates in the SED fitting analysis, results in $\sim0.1$\,dex variance in the estimated values. The values for the host galaxy properties derived using the templates shown in Table \ref{table_cigale} and used in our clustering analysis are within these fluctuations.


\begin{figure}
\includegraphics[height=0.9\columnwidth]{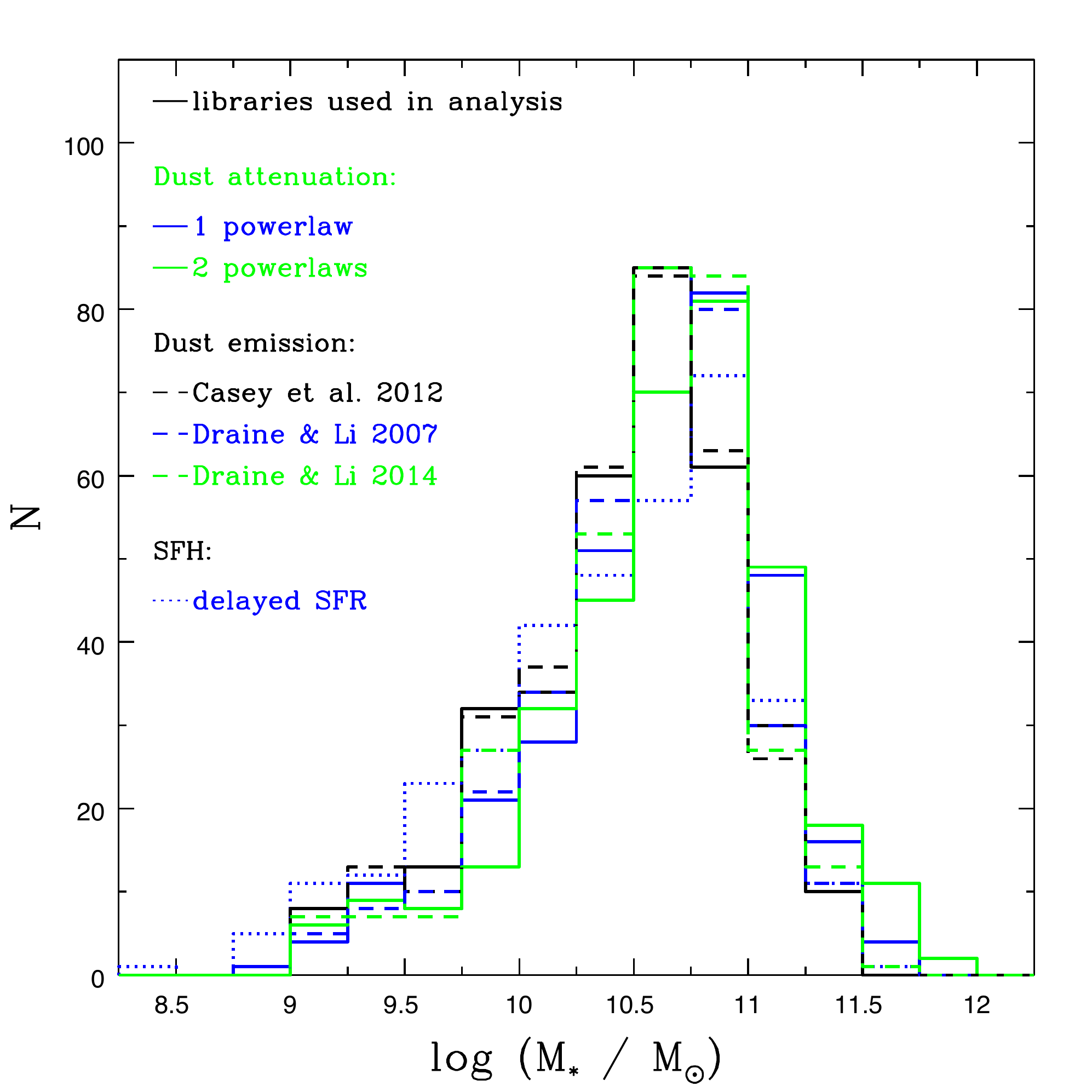}
\includegraphics[height=0.9\columnwidth]{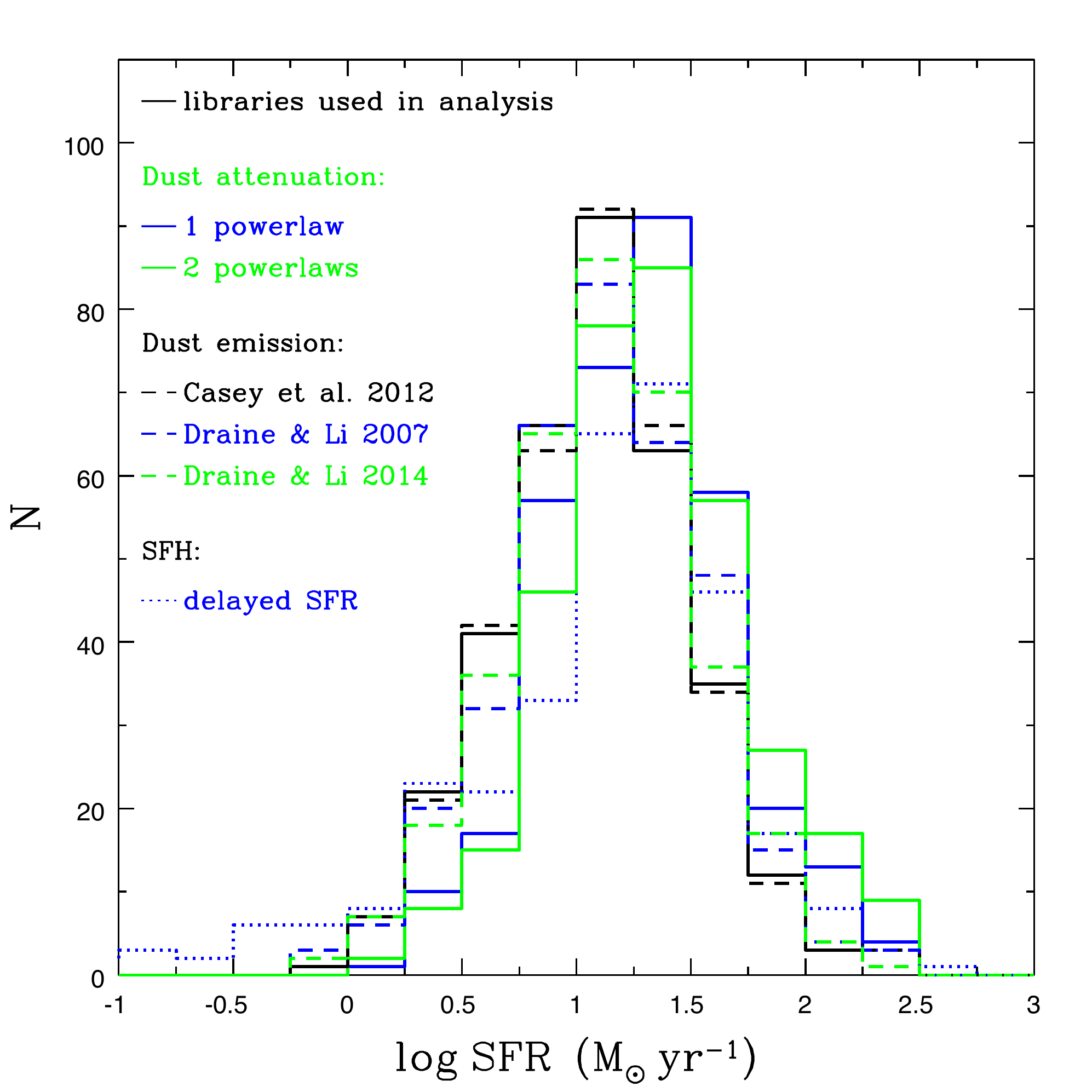}
\includegraphics[height=0.9\columnwidth]{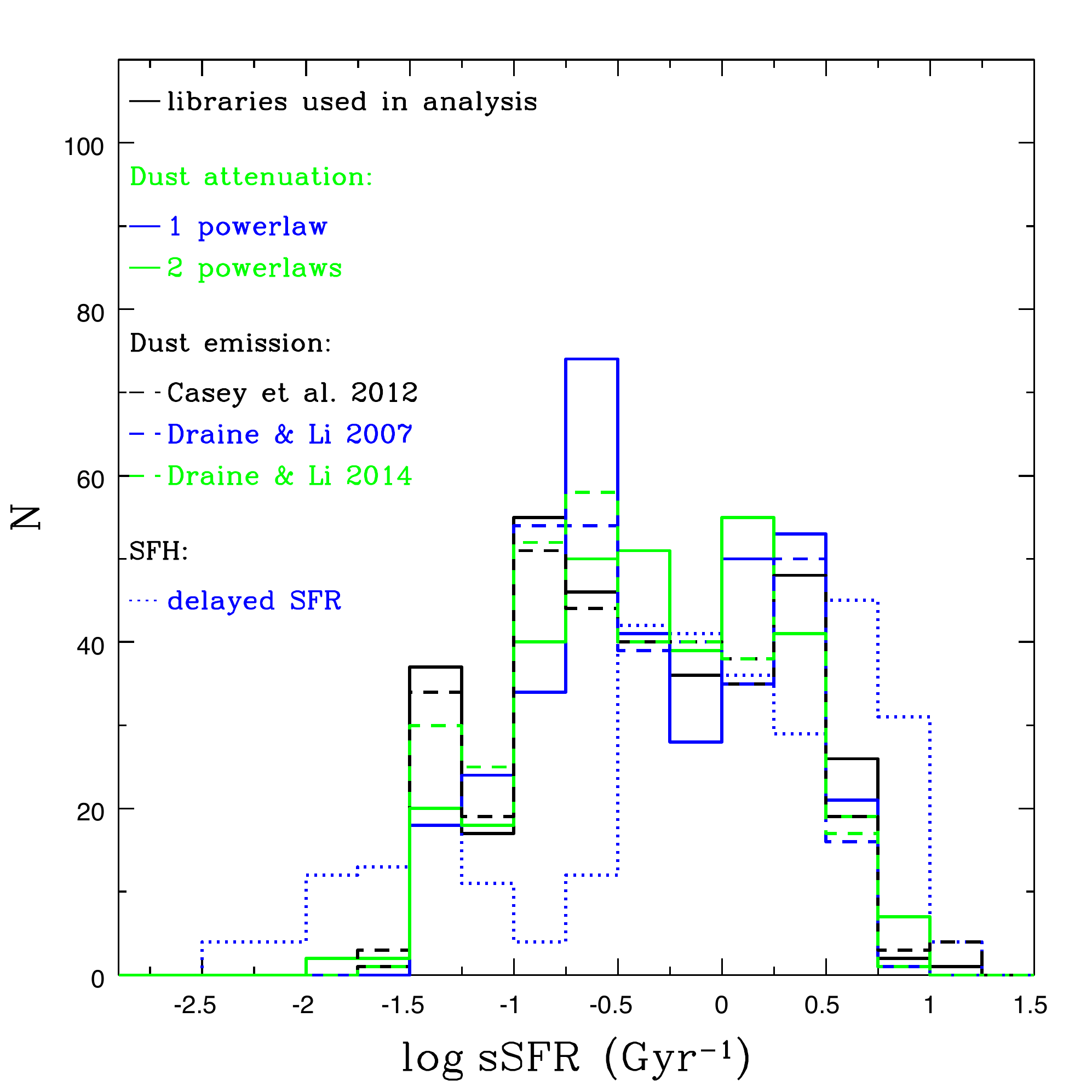}
\caption{From top to bottom: M$_*$, SFR and  sSFR distributions of the AGN sample, using different libraries for the SED fitting. Draine \& Li (2014) refers to the implementation of the updated Draine \& Li (2007) models.}
\label{fig_diff_modules}
\end{figure}

\section{VIPERS galaxy clustering: PDR2 vs. PDR1}
\label{append_vipers}

As mentioned in Section \ref{results} and presented in Table \ref{table:clustering_values} the galaxy autocorrelation function estimated in this work appears higher compared to the measurements of \cite{Mountrichas2016} (Fig. \ref{clustering_measurements}). This may be attributed to the differences in the galaxy samples used in the two studies. These are: (a) the larger area ($\approx 2\times$) covered by PDR2 (this study) compared to PDR1 \citep{Mountrichas2016} and the improved estimations of the SSR and TSR parameters in PDR2 galaxy sample \citep[see][]{Scodeggio2016}.

To examine whether the cosmic variance affects the galaxy measurements and to which degree, we measure the galaxy autocorrelation function using galaxies from the PDR2 that are not included in the PDR1 VIPERS catalogue. The results are presented in Fig. \ref{fig_tsr} (red squares). The corresponding galaxy bias is b$=1.32^{+0.05}_{-0.04}$ compared to  b$=1.22^{+0.03}_{-0.03}$ using the PDR1 galaxy sample (red triangles).

To test the effect of the improved TSR and SSR parameters on our galaxy clustering measurements, we measure the galaxy autocorrelation function, using the PDR1 galaxy sample with the improved TSR, SSR parameters (blue circles in Fig. \ref{fig_tsr}). The corresponding bias value is b$=1.26^{+0.03}_{-0.04}$.

Based on the above calculations, most of the increased clustering signal of the PDR2 galaxy dataset compared to the PDR1 can be attributed  to the increased area covered by the PDR2 sample (cosmic variance) and to a lesser degree to the improved estimations of the TSR and SSR parameters.

\begin{figure}
\begin{center}
\includegraphics[height=1.1\columnwidth]{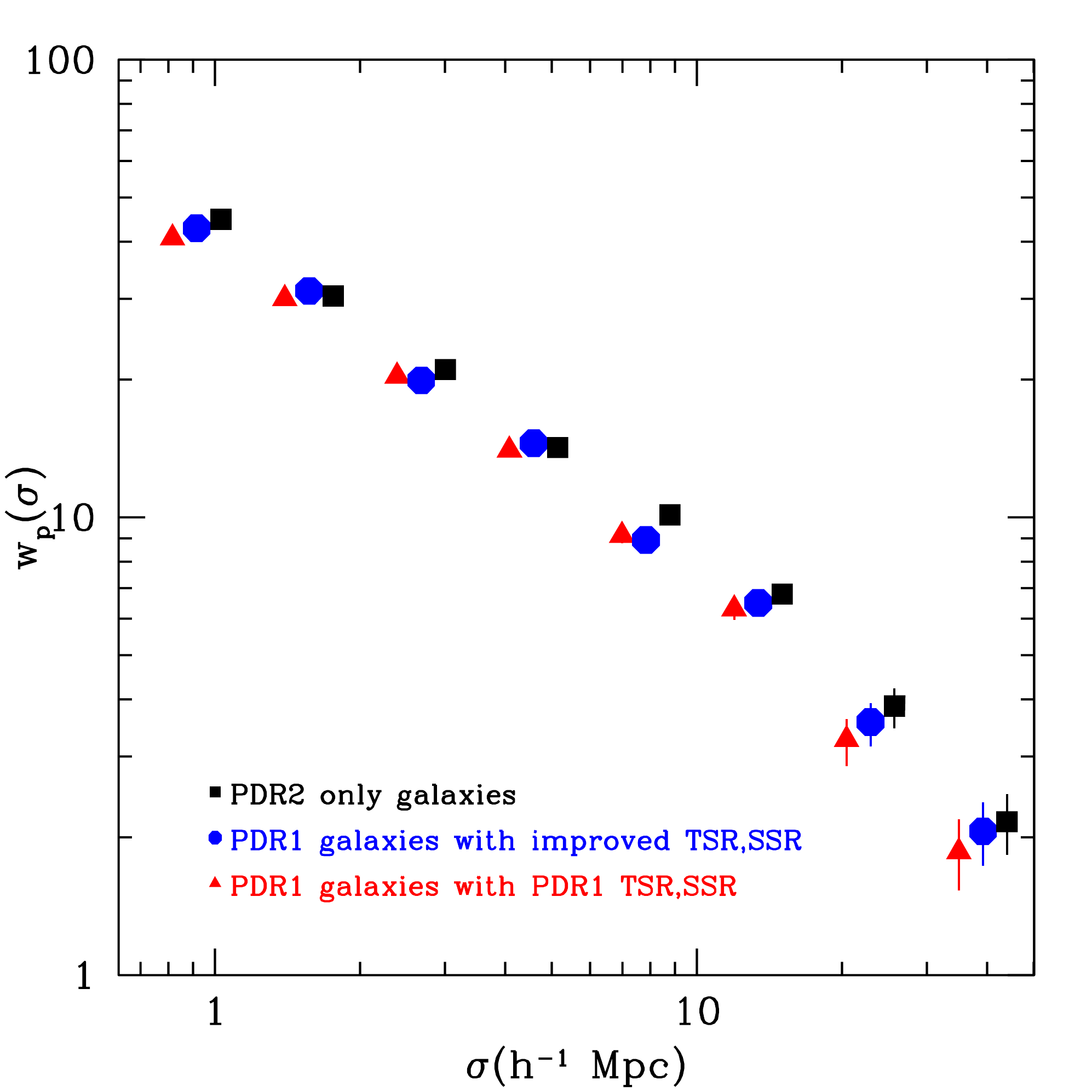}
\end{center} 
\caption{Galaxy autocorrelation function using the PDR1 of the VIPERS catalogue (triangles), the updated TSR, SSR estimations (circles) and galaxies only within the PDR2 dataset (squares).Triangles and squares are offset in the horizontal direction by $\log=-0.05$ and $\log=+0.05$ for clarity.}
\label{fig_tsr}
\end{figure}

\end{document}